\documentclass[pdflatex,sn-apa]{sn-jnl} 

\usepackage{graphicx}%
\usepackage{multirow}%
\usepackage{amsmath,amssymb,amsfonts}%
\usepackage{amsthm}%
\usepackage{mathrsfs}%
\usepackage[title]{appendix}%
\usepackage{xcolor}%
\usepackage{textcomp}%
\usepackage{manyfoot}%
\usepackage{booktabs}%
\usepackage{algorithm}%
\usepackage{algorithmicx}%
\usepackage{algpseudocode}%
\usepackage{listings}%

\theoremstyle{thmstyleone}%
\theoremstyle{thmstyletwo}%

\theoremstyle{thmstylethree}%

\begin{document}

\title[Revisiting Global Income Convergence in the 21st Century]{%
Revisiting Global Income Convergence in the 21st Century}

\author*[1]{\fnm{Bipul} \sur{Verma}}\email{verma176@umn.edu}

\affil*[1]{\orgdiv{Department of Economics}, \orgname{University of Minnesota},  \orgaddress{\city{Minneapolis}, \postcode{55454}, \state{Minnesota}, \country{United States }}}


\abstract{Recent research has documented a reversal from divergence to convergence in income levels between rich and poor countries after 2000. This paper employs a growth accounting framework to investigate the proximate sources of convergence over the period 1980-2019. I find that while output levels began to converge only after 2000, capital (physical and human) had already been converging during 1980-2000. The divergence in total factor productivity (TFP) during this earlier period offset the gains from capital convergence, resulting in little overall income convergence. After 2000, capital maintained its pattern of convergence, and, unlike in the earlier period, TFP also began to converge.  Quantitatively, more than half of the convergence between 2000 and 2019—and nearly all the convergence between 1980 and 2019 outside Sub-Saharan Africa—can be attributed to the convergence in capital rather than to improvements in TFP. These findings highlight that factor accumulation has played a much larger role in driving long-run convergence dynamics than previously documented in the literature.}




\maketitle

\section{Introduction}\label{sec_intro}
The traditional literature on economic growth has long debated whether poorer countries' incomes converge toward those of richer countries. Empirical studies, particularly those prior to the early 2000s, generally found limited or no support for unconditional convergence \citep{barro1992convergence, pritchett1997divergence}. However, recent studies by \cite{patel2021new} and \cite{kremer2022converging} challenge this view, documenting a reversal from divergence to convergence in recent decades.  This documented reversal prompts two critical yet relatively unexplored questions: first, what factors can explain this turnaround in convergence dynamics; and second, which proximate determinants have contributed significantly to the observed income convergence?

I address these questions by first replicating the empirical finding of convergence in recent decades, thereby establishing a consistent empirical basis for further analysis. I then examine the proximate drivers of this shift using a growth accounting framework that decomposes income growth into contributions from total factor productivity (TFP), physical capital, and human capital. In doing so, I explicitly account for heterogeneity in capital income shares across countries, a feature that standard growth accounting exercises often abstract from. This approach allows for a more accurate quantification of the role played by each factor in driving convergence, and highlights how their relative contributions have changed over time.

I find that the reversal from divergence to convergence can be attributed to two main factors: first, total factor productivity (TFP) shifted from a diverging pattern before 2000 to convergence thereafter; and second, physical and human capital inputs continued to converge consistently. While TFP transitioned from divergence to convergence post-2000, its contribution to long-run income convergence remains secondary. I show that over the period 2000-2019, more than half of the observed income convergence among countries outside Sub-Saharan Africa (SSA) arises from convergence in physical and human capital. More strikingly, over the longer timeframe of 1980-2019, income convergence among countries outside SSA occurs \emph{despite}  divergence in total factor productivity (TFP). The entirety of the long-run convergence in output levels for these countries is driven solely by narrowing gaps in physical and human capital accumulation. Additionally, I find that for poorer countries within Sub-Saharan Africa (SSA), there has been no convergence in the capital-output ratio since 2000. Notably, while the extent of total factor productivity (TFP) convergence between rich economies and poor countries outside SSA mirrors that observed between rich economies and poor countries in SSA, the markedly lower degree of output convergence in the latter is entirely accounted for by the absence of capital convergence. This pattern suggests that the lack of capital convergence is an important factor contributing to the region's relative economic stagnation.

These findings provide a revised understanding of the drivers behind global income dynamics. Prior growth accounting and development accounting exercises consistently highlighted differences in total factor productivity (TFP) as the central factor explaining persistent income gaps and divergence across countries \citep{klenow1997neoclassical, easterly2001have, jones2016facts}. While TFP indeed remains the dominant explanatory factor in the divergence era preceding 2000, its relative importance diminishes significantly thereafter. Specifically, for the period following 2000, and even more clearly over the longer span from 1980 to 2019, the observed convergence in global income levels is primarily driven by convergence in physical and human capital accumulation rather than TFP.  The differences between these findings and the prior literature primarily reflect two key factors: first, this paper explicitly analyzes the more recent period of convergence post-2000, whereas earlier studies predominantly considered periods of divergence; second, the analysis incorporates heterogeneity in capital income shares across countries and over time. Assuming a constant capital income share  underestimates the contribution of factor inputs, especially in the period following 2000. This paper also contributes to the literature on economic growth in Sub-Saharan Africa (SSA) by identifying the non-convergence of capital accumulation as a critical factor underlying the region's persistent income stagnation. 

The findings in this paper have important theoretical and policy implications. Economic growth models should incorporate heterogeneity in capital income shares and place greater emphasis on growth driven by the accumulation of physical and human capital. Given that these factors have been the primary drivers of income convergence, policymakers should prioritize investments in infrastructure, machinery, education, and healthcare—especially in regions such as Sub-Saharan Africa where growth has been slow.

\section{Data}\label{sec_data}
This section summarizes the data used in the current study. Data for real GDP adjusted for purchasing power parity (PPP) ($rgdpo$), population ($pop$), human capital ($hc$), and labor share of income ($labsh$) for all countries come from the Penn World Tables (PWT10.01).\footnote{See \cite{feenstra2015next} for details.} Regional classifications are based on the World Bank's classification of economies. I calculate the capital-output ratio as the ratio of real capital stock (\textit{rnna}) to real GDP (\textit{rgdpna}). I use per capita GDP, adjusted to 2017 PPP dollars, as the measure of income ($rgdpo/pop$). The selection of output side real gdp (rgdpo) over alternative Penn World Table output measures follows the recommendations of \cite{feenstra2015next} for comparing productive capacities across countries and over time. 

All samples excludes countries with small populations and those whose GDP is heavily dependent on oil rents, which is standard practice in the literature.\footnote{Specifically, I exclude countries whose population never exceeded 0.2 million at any point in time. I also exclude countries where oil rents accounted for more than 50\% of GDP at any time. The capital measure in the PWT includes fixed reproducible capital but excludes natural resource capital, such as land, inventories, subsoil assets, or intangible capital. This difference is particularly significant for oil-rich countries, which is why they are excluded from the main analysis. \cite{kremer2022converging} also excludes small countries, and oil rich countries in their analysis.} I use a sample of 84 economies outside Sub-Saharan Africa, for which the PWT reports human capital, physical capital, and capital share of income to conduct the main growth decomposition exercise.\footnote{For variance measures, I further exclude Venezuela. Venezuela is excluded for statistical reasons, as it experienced an economic crisis beginning in 2000, with current GDP per capita lower than its 1960 level. This makes Venezuela an outlier with a large influence on variance measures. The time series of Venezuela's GDP per capita is plotted in the supplementary appendix for reference.} 

\section{Income Convergence}\label{sec_bet_con}
I replicate the empirical framework of \cite{patel2021new} and \cite{kremer2022converging} to investigate whether, in the post-2000 period, economies with lower initial income per capita have tended to grow more rapidly than those with higher initial incomes. This pattern-commonly referred to as unconditional (or $\beta$-) convergence-implies that poorer countries catch up to richer ones over time. To quantify the speed of convergence between dates $t$ and $T$, I estimate the following specification:
\begin{equation}\label{eqn_reg_beta}
\frac{1}{s} \ln \Big(\frac{y_{i, t+s}}{y_{i, t}} \Big) = \beta_0 - \Big( \frac{1-\exp^{\beta s}}{s} \Big)\ln(y_{i, t}) + \epsilon_{i, t+s}.
\end{equation}
Here, $y_{i,t}$ refers to the income of country $i$ in year $t$, $y_{i,t+s}$ refers to the income of country $i$ in year $t+s$, and $\epsilon_{i,t+s}$ captures the shocks to the annualized growth rate between time periods $t$ and $t+s$. I assume that these shocks are  independently and identically distributed (iid) with a normal distribution. I estimatate specification \ref{eqn_reg_beta} using non-linear least square with heteroskedaticity-robust standard errors. I estimate separate regressions for the periods 1980-2000 and 2000-2019. In this specification, $\beta$ is identified using cross-sectional variation in annual growth rates across countries initially at different income levels. Additionally, note that the regression in equation \ref{eqn_reg_beta} includes no country-level controls, meaning that the $\beta$ identified here reflects the rate of unconditional convergence. 

\begin{table}[htbp]\centering
\caption{Beta Convergence for 1980-2000 and 2000-2019.}
\begin{tabular}{l*{4}{c}}
\hline\hline
            &\multicolumn{2}{c}{All Countries} &\multicolumn{2}{c}{Outside SSA} \\
            &\multicolumn{1}{c}{(1)}&\multicolumn{1}{c}{(2)}&\multicolumn{1}{c}{(3)}&\multicolumn{1}{c}{(4)}\\
            &\multicolumn{1}{c}{1980-2000}&\multicolumn{1}{c}{2000-2019}&\multicolumn{1}{c}{1980-2000}&\multicolumn{1}{c}{2000-2019}\\
\hline
$\beta$       &       0.0047**&      -0.0062***&      0.0006         &      -0.0150***\\
            &     (0.001)         &     (0.002)         &     (0.001)         &     (0.003)         \\
\hline
N & 131 & 155 & 86 & 115 \\ 
\hline\hline
\multicolumn{5}{l}{\footnotesize Standard errors in parentheses}\\
\multicolumn{5}{l}{\footnotesize * \(p<0.10\), ** \(p<0.05\), *** \(p<0.01\)}\\
\end{tabular}
\label{tab_beta_est}
\end{table}

Table \ref{tab_beta_est} presents estimates of the rate of $\beta$-convergence. Columns (1) and (2) report results for the full balanced panel, whereas columns (3) and (4) restrict the sample by excluding Sub-Saharan African economies. Over the 1980-2000 period, the coefficient in column (1) is positive and statistically significant, indicating that wealthier countries grew faster than poorer ones and thereby increasing income divergence. The coefficient in column (3) does not differ significantly from zero, showing that even when Sub-Saharan Africa is excluded, there was no convergence over the 1980-2000 period.

In contrast, the rate of convergence has seen a remarkable reversal in the past two decades. In the full balanced panel, initially poorer economies have closed their income gap at an annual rate of 0.62 percent. When Sub-Saharan African countries are excluded, the catch-up rate rises to 1.5 percent per year. By comparison, \cite{patel2021new} report a 0.425 percent convergence rate for 2000-2019 in their full sample and 1.35 percent outside Sub-Saharan Africa, while \cite{kremer2022converging} document a 0.7 percent rate for 2005–2015 (base year 2005) in a broad country panel. The somewhat higher coefficient estimated here reflects differences in sample composition and output measurement: \cite{patel2021new} employ expenditure-side real GDP (rgdpe) and omit countries with populations below one million, whereas this analysis uses output-side real GDP (rgdpo) and excludes economies with fewer than 0.2 million inhabitants, in line with \cite{kremer2022converging}. Table \ref{tab:beta_comparision} presents the convergence coefficients across alternative Penn World Table output series for both samples.

The \textit{unconditional} convergence coefficient for the post 2000  period outside SSA is also close to the rate of \textit{conditional} convergence found in recent literature. \cite{barro2015convergence} estimates the rate of \textit{conditional} convergence to be 1.7\% regressing 5-year growth rates on lagged income level and country level-time varying controls.  The \textit{unconditional} convergence rate of 1.5\% observed here  for countries outside SSA is substantial in comparison. This rate of $\beta-$convergence rate of 1.5\% implies a half-life of 53 years, meaning that, starting from 2020, poor countries outside Sub-Saharan Africa would close half the income gap with developed countries by 2073, assuming all countries share the same steady-state income level. \footnote{The half-life is calculated as $\tau = -\ln(2)/\ln\Big( 1- \frac{1-\exp^{\beta s}}{s} \Big)$.}

\begin{table}[t]
\centering
\caption{Income Differences across years (outside SSA)}\label{tab_inc_diff}
\begin{tabular}{r r r r r r}  
\hline
Year & P90/P10 & P90/P50 & P50/P10 & Var(log(GDPpc)) & Income Ratio\\
\hline
\textbf{Outside SSA} & & & & \\
1980 & 15.75 & 3.22 & 4.89 & 1.08 & 35.29\\
1990 & 19.81 & 3.49 & 5.67 & 1.17 & 37.08\\
2000 & 17.10 & 3.87 & 4.42 & 1.20 & 38.30\\
2010 & 11.22 & 2.92 & 3.85 & 0.91 & 34.65\\
2019 & 9.97 & 2.73 & 3.64 & 0.81 & 24.78\\
\hline
\textbf{All Countries} & & & & \\
1980 & 18.56 & 5.30 & 3.50 & 1.15 & 18.49\\

1990 & 24.95 & 5.55 & 4.50 & 1.37 & 23.26\\

2000 & 31.95 & 6.88 & 4.64 & 1.62 & 29.53\\

2010 & 27.63 & 4.51 & 6.12 & 1.54 & 30.71\\

2019 & 25.06 & 4.41 & 5.69 & 1.47 & 27.17\\
\hline 
\hline 
\end{tabular}
\end{table}

Beta-convergence is necessary but not sufficient for sigma-convergence \citep{young2008sigma, barro1992convergence}.\footnote{Sigma convergence refers to a decline over time in the cross-sectional dispersion of per-capita income across economies.} Consequently, even if initially poorer economies grow more rapidly than their richer counterparts, the cross-sectional dispersion of per-capita income need not decline. Table \ref{tab_inc_diff} reports five measures of income dispersion-namely the P90/P10, P90/P50 and P50/P10 percentile ratios of GDP per capita; the variance of log GDP per capita; and the ratio of the top five to bottom five incomes-for each decade from 1980 to 2019. The table enables a clear assessment of whether-and when-sigma-convergence accompanied the episodes of unconditional convergence identified earlier.

For countries outside SSA, income dispersion widened during the 1980s—by 1990 the P90/P10 ratio had risen to nearly 20 and the variance of log GDP per capita to  above 1.1. Since 2000, every measure has fallen steadily: by 2019 the P90/P10 ratio had roughly halved to below 10, and the variance of log GDP per capita declined to  0.8, reaching its lowest level in four decades. Hence, after 2000, initially poorer non-SSA economies not only grew faster than their richer counterparts but also saw a marked reduction in overall income dispersion—demonstrating both beta- and sigma-convergence in that subsample.

For the full country sample, dispersion rose sharply through the 1980s and 1990s, with the P90/P10 ratio climbing from 18.6 in 1980 to nearly 32.0 in 2000 before easing to 25.1 by 2019. Likewise, the variance of log GDP per capita peaked at 1.62 in 2000 and then fell modestly to 1.47 by 2019. By contrast, the P50/P10 ratio rose steadily from 3.50 in 1980 to 5.69 in 2019, signaling a widening gap between middle-income and poorest countries. These trends imply only limited sigma-convergence: although top-to-bottom dispersion declined somewhat after 2000, overall income inequality in 2019 remains well above its 1980 level.

Table \ref{tab_inc_diff} demonstrates that, although economies outside Sub-Saharan African achieved both beta- and sigma-convergence after 2000, many Sub-Saharan African countries have continued to stagnate. Since most of the observed convergence-and the most reliable data-resides outside SSA \citep{young2012african}, the subsequent analysis focuses on non-SSA economies.  Full-sample results appear in the Appendix.

\section{Accounting for Decline in Income Dispersion}\label{sec_dev_acc}
In this section, I perform an accounting exercise to attribute the decline in income dispersion to reductions in TFP differences and input dispersion. Unlike many standard approaches in the literature, which assume a uniform capital share parameter of one-third, this accounting exercise based on percentile ratios allows the capital share of income to vary continuously with the income percentile, as observed in PWT10.01.  The accounting exercise using variance as a measure of income dispersion is presented in the appendix.

Let $p$ denote the income percentile. Suppose per-worker output is generated by a constant-returns-to-scale technology
\[
  y(p)=F\bigl(A(p),k(p),h(p)\bigr),
\]
where $A(p)$ is Hicks-neutral productivity.\footnote{It is common to interpret $A(p)$ as total factor productivity (TFP), but more broadly it includes both observable and unobservable factors that affect output beyond the inputs. As argued in \cite*{klenow1997neoclassical}, under firm optimization, equalization of the marginal product of capital with the rental rate implies that TFP growth mechanically raises the capital stock under a constant rental rate; to separate TFP and capital-accumulation effects in output growth, the production function is written in intensive form.} 
Then the difference in log-output between percentiles $p_{0}$ and $p_{n}$ admits the decomposition
\begin{equation}\label{eqn:main_decomp}
  \ln y(p_{n}) - \ln y(p_{0})
  =\int_{p_{0}}^{p_{n}}\Bigl[
    \frac{1}{1 - \alpha(p)}\,\frac{d\ln A(p)}{dp}
   +\frac{\alpha(p)}{1 - \alpha(p)}\,\frac{d\ln\bigl(k/y\bigr)(p)}{dp}
   +\frac{d\ln h(p)}{dp}
  \Bigr]\,dp.
\end{equation}
where $\alpha(p)$ denotes the capital share of income at percentile $p$. Here, the first term captures the contribution of TFP, the second term captures the contribution of capital-output ratio, and the third term captures the contribution of human capital.  Appendix \ref{app_maths} provides the full derivations of each elasticity under CRS.

I report the results from the decomposition exercise for the percentile ratio-P90/P10- covering countries outside Sub-Saharan Africa for the time periods: 1980-2000, 2000-2019. The decomposition results for P90/P50, P50/P10, as well as for full sample of countries are reported in Appendix.\footnote{I refer P90, P50, P10 as rich, middle, poor economies respectively.}

\subsection*{Decomposition Results: Convergence outside SSA}
\begin{figure}[htbp]
\centering
\includegraphics[width=\linewidth]{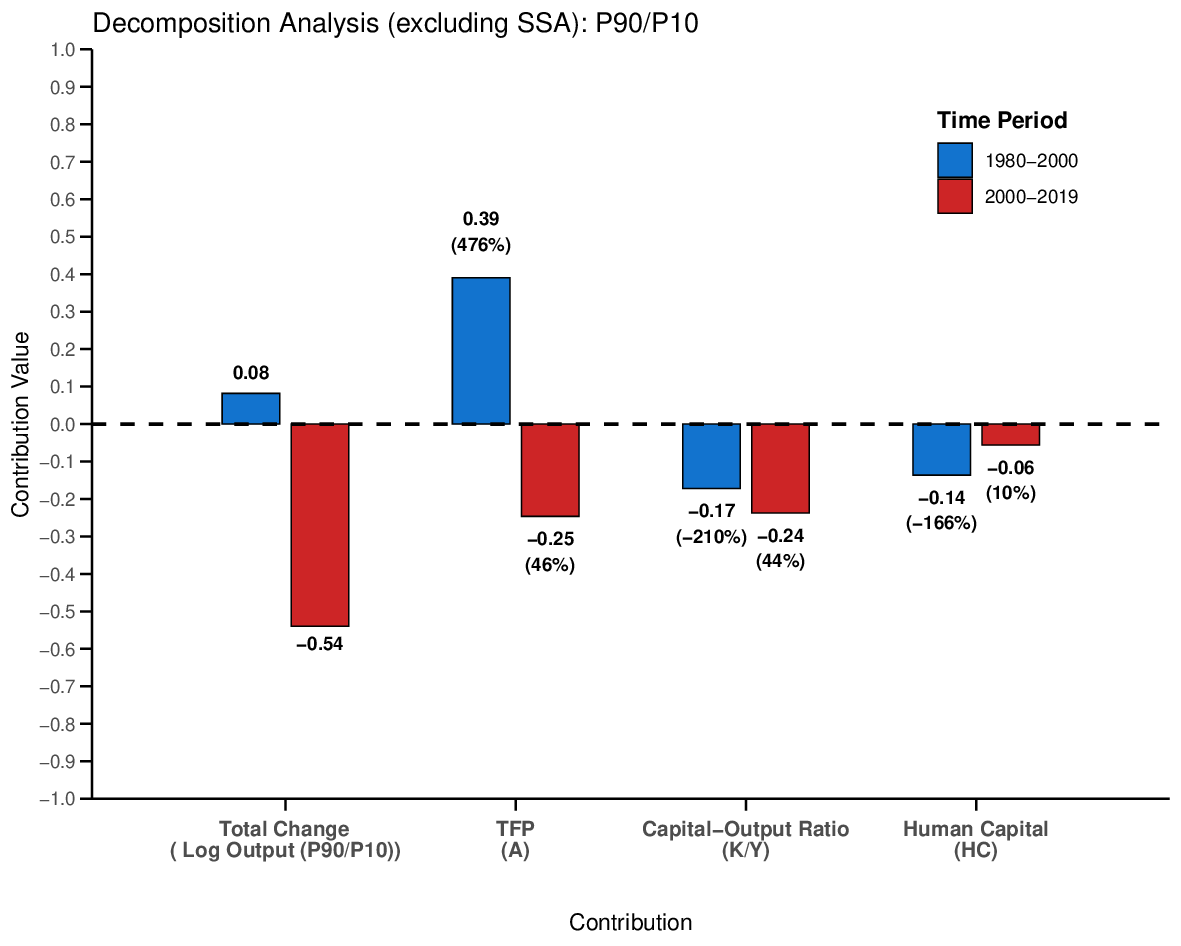}
\caption{Decomposition Exercise (excluding SSA): P90/P10}
\label{fig:decomposition_exercise_excluding_ssa_p90_p10}
\end{figure}

Figure \ref{fig:decomposition_exercise_excluding_ssa_p90_p10} illustrates the decomposition of the P90/P10 income ratio over two distinct time periods: 1980-2000 (blue bars), and 2000-2019 (red bars). The vertical axis shows the change in values for each period, while the horizontal axis shows the overall change in the log P90/P10 ratio followed by the contributions of TFP, the capital-output ratio, and human capital to that change, as derived in equation \eqref{eqn:main_decomp}.

Convergence in both physical and human capital inputs was the principal force shaping the evolution of the P90/P10 income ratio. Between 1980 and 2000, the P90/P10 income ratio increased by 8 percentage points, reflecting a period of divergence between rich and poor countries. Before 2000, TFP divergence alone would have driven the ratio up by roughly 0.39, but simultaneous convergence in physical capital and human capital reduced the ratio by about 0.17 and 0.14, respectively. 

From 2000 to 2019 the pattern reversed.  The log P90/P10 ratio fell by \(0.54\).  Convergence in the capital-output ratio contributed \(0.24\) of this reduction (about 44\%), while human-capital convergence contributed \(0.06\) (10\%).  During this interval TFP shifted from divergence to convergence, contributing \(0.25\) (46\%) to  overall convergence. The shift in total factor productivity (TFP) dynamics---from divergence before 2000 to convergence thereafter---played a pivotal role in driving post-2000 income convergence. In the absence of TFP convergence, the extent of overall convergence during this period would have been minimal. Nevertheless, in quantitative terms, the contribution of TFP to post-2000 convergence remains smaller than that of input accumulation, accounting for 46\% compared to 54\%. This contrast is even more pronounced when examining the earlier period from 1980 to 2000.

Note that the overall change in the P90/P10 ratio from 1980 to 2019 is the sum of the changes over 1980-2000 and 2000-2019, as expressed by the additive property of logarithms. Over the full horizon 1980-2019 the log P90/P10 ratio declined by \(0.46\).  Cumulatively, the capital-output ratio accounted for \(0.41\) of the decline (89\%) and human capital for \(0.19\) (41\%), whereas persistent TFP divergence added back roughly \(0.14\).  Despite the continued divergence in TFP, there was substantial convergence in income over the period 1980-2019. This implies that convergence in physical and human capital inputs fully accounts for the observed income convergence during this period.

The Appendix reports analogous decompositions for the P90/P50 and P50/P10 ratios, respectively.  For the P90/P50 ratio input convergence accounts for 66\% of the decline observed in 2000-2019 and explains the entire convergence over the full 1980-2019 horizon.  In the case of the P50/P10 ratio inputs contribute less—32\% of the convergence during 2000-2019 and 70\% over 1980-2019.  In both decompositions the behavior of total factor productivity mirrors that documented for the P90/P10 ratio, diverging in 1980-2000 and converging in 2000-2019.  Taken together, the evidence indicates that poorer economies outside Sub-Saharan Africa are catching up not only with rich but also with middle-income countries. This long-run convergence---particularly after 2000, and more importantly over the full period from 1980 to 2019---is driven primarily by the convergence of physical and human capital inputs.

\subsection*{Decomposition Results: Stagnation of SSA}
\begin{figure}[htbp]
\centering
\includegraphics[width=\linewidth]{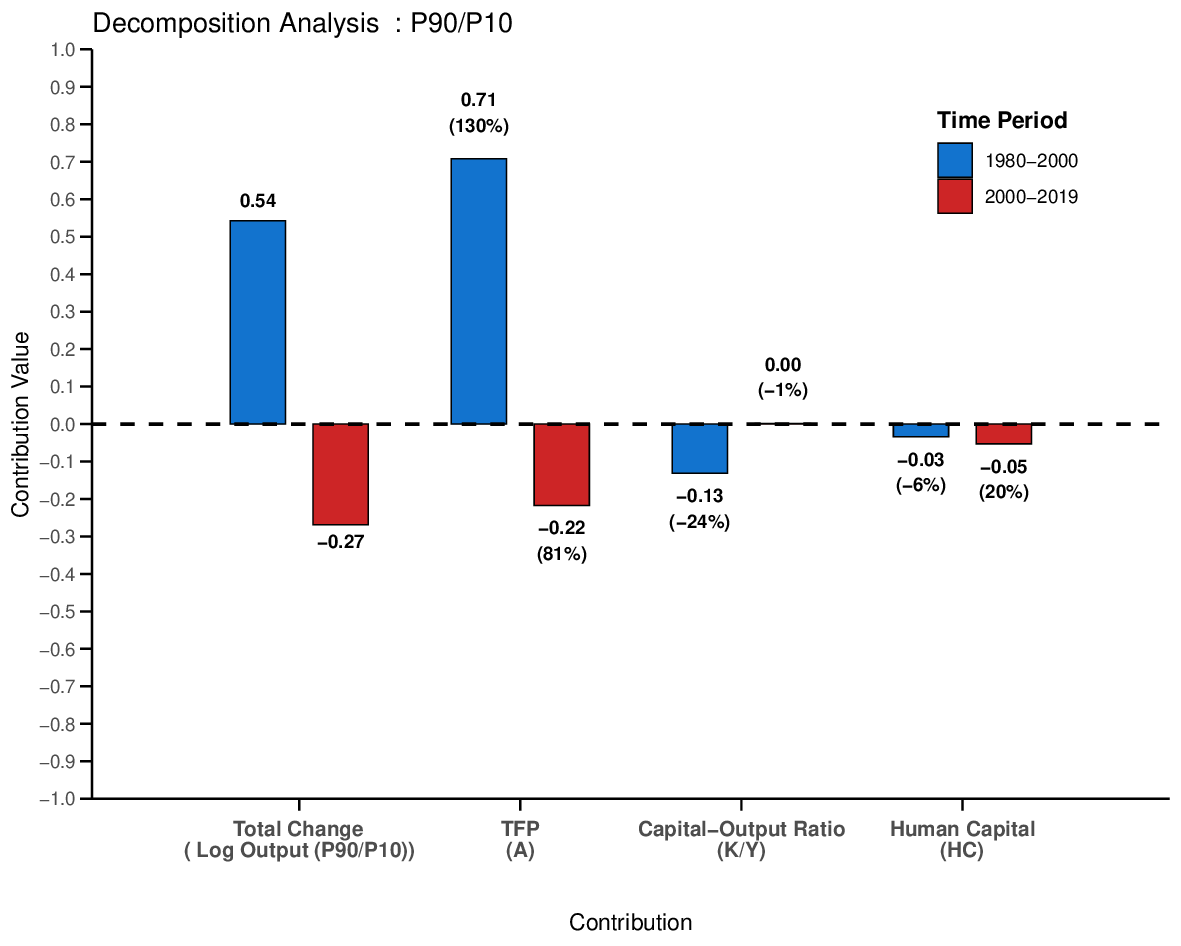}
\caption{Decomposition Exercise (full sample): P90/P10}
\label{fig:decomposition_exercise_full_sample_p90_p10}
\end{figure}

Figure \ref{fig:decomposition_exercise_full_sample_p90_p10} presents the decomposition results for the full sample of countries, including those in Sub-Saharan Africa (SSA). It is important to note that, as of 2000, economies below the P10 income level were predominantly from Sub-Saharan Africa. Therefore, this decomposition provides insights into the stagnation of poor economies in SSA relative to those outside the region.   Similar to poor countries outside SSA, economies in SSA experienced divergence in income during 1980-2000 and convergence toward richer countries after 2000. However, the degree of divergence before 2000 was substantially larger within SSA (0.54 compared to 0.08), and the extent of convergence since 2000 was correspondingly much smaller (0.27 compared to 0.54). The larger divergence before 2000 for SSA economies is predominantly driven by more pronounced divergence in total factor productivity (TFP): 0.71 compared to 0.39 for non-SSA countries. Lower convergence in human capital also contributed: 0.03 for SSA compared to 0.14 outside SSA—but its role was relatively modest in comparison to TFP. Convergence in physical capital, by contrast, was broadly similar across the two groups, at 0.13 for SSA and 0.17 for non-SSA economies.

In the post-2000 period, convergence in TFP (0.22 compared to 0.25) and in human capital (0.05 compared to 0.06) were of similar magnitudes between SSA and non-SSA poor economies. The key distinction lies in physical capital: the absence of convergence in physical capital among SSA economies entirely accounts for the lower extent of output convergence relative to poor countries outside the region.

\begin{figure}[htbp]
  \centering
      \includegraphics[width=\linewidth]{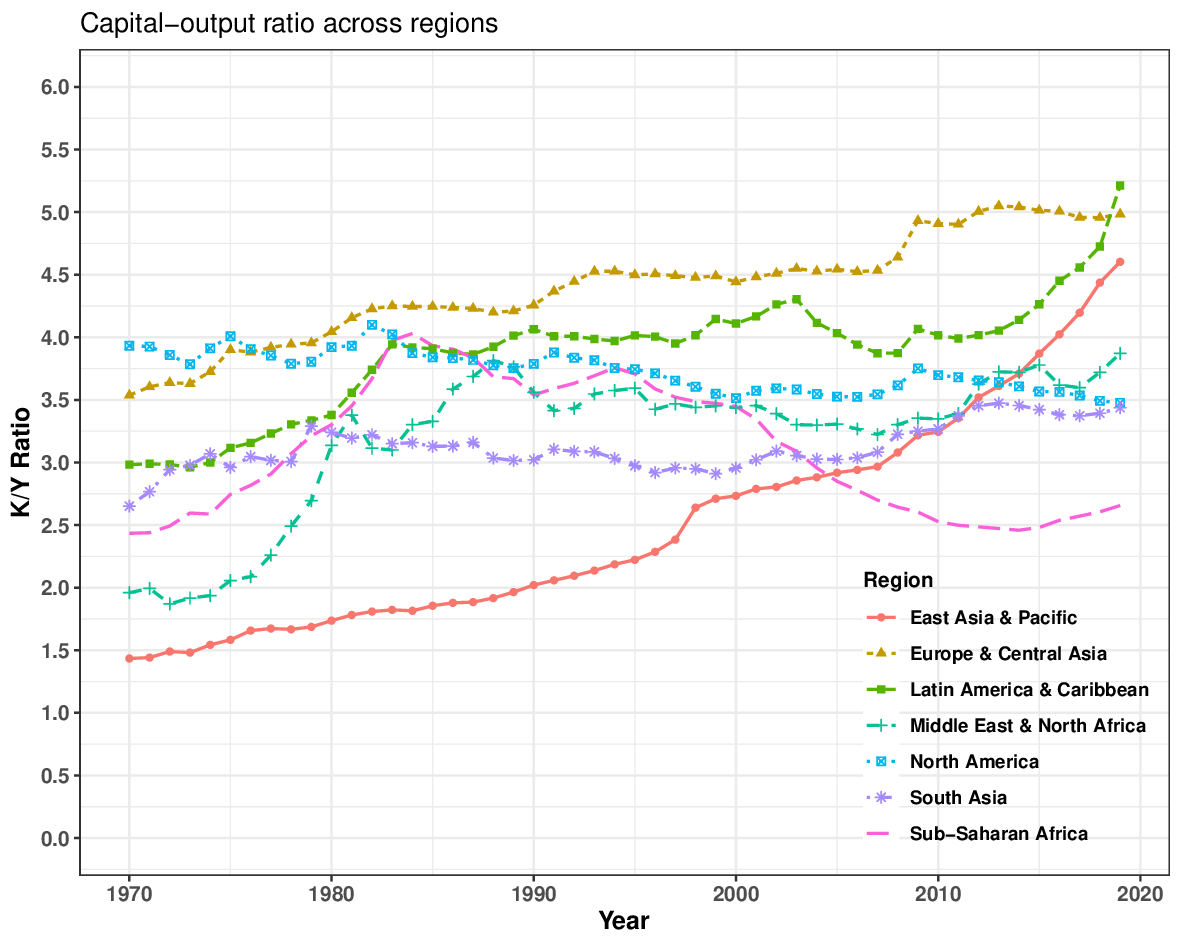}
  \caption{Capital-Output Ratio across regions}
  \label{fig_cap_out_regions}
  \end{figure}

Unlike poor economies outside SSA, SSA did not exhibit any meaningful convergence in the capital-output ratio post-2000, and only negligible convergence in human capital. While TFP did converge post-2000, the lack of convergence in inputs meant that the forces of convergence were insufficient to offset the TFP divergence that occurred prior to 2000. As a result, between 1980 and 2019, SSA economies fell further behind the rich economies due to TFP divergence in the early period and the absence of input convergence across the full time span. Further evidence of the lack of input convergence, particularly in capital, is presented in Figure \ref{fig_cap_out_regions}, which illustrates the mean capital-output ratio, weighted by population size, across different regions. While the capital-output ratio has continued to rise in all regions outside Sub-Saharan Africa, SSA is the only region to exhibit a decline in its capital-output ratio post-2000. This decline likely reflects the region's struggles with capital accumulation—challenges that may be linked to political instability, economic uncertainty, and capital flight, as discussed in \cite{ndikumana2011africa}.

\subsection*{Decomposition with constant capital income share}

\begin{figure}[htbp]
  \centering
  \includegraphics[width=\linewidth]{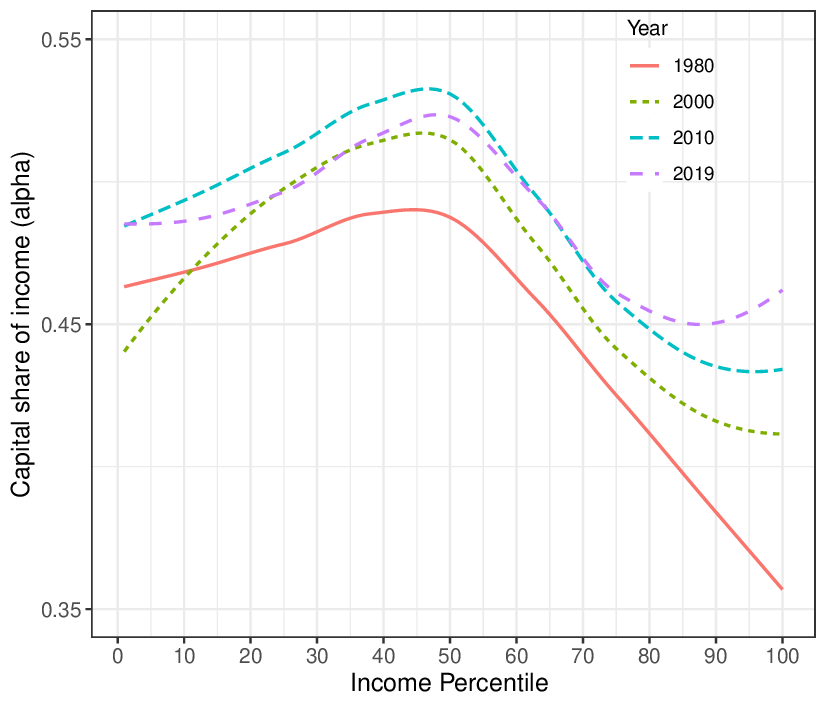}
  \caption{ Variation of alpha over income percentile.}
  \label{fig_alpha_ptile}
\end{figure}

\begin{figure}[htbp]
\centering
\includegraphics[width=\linewidth]{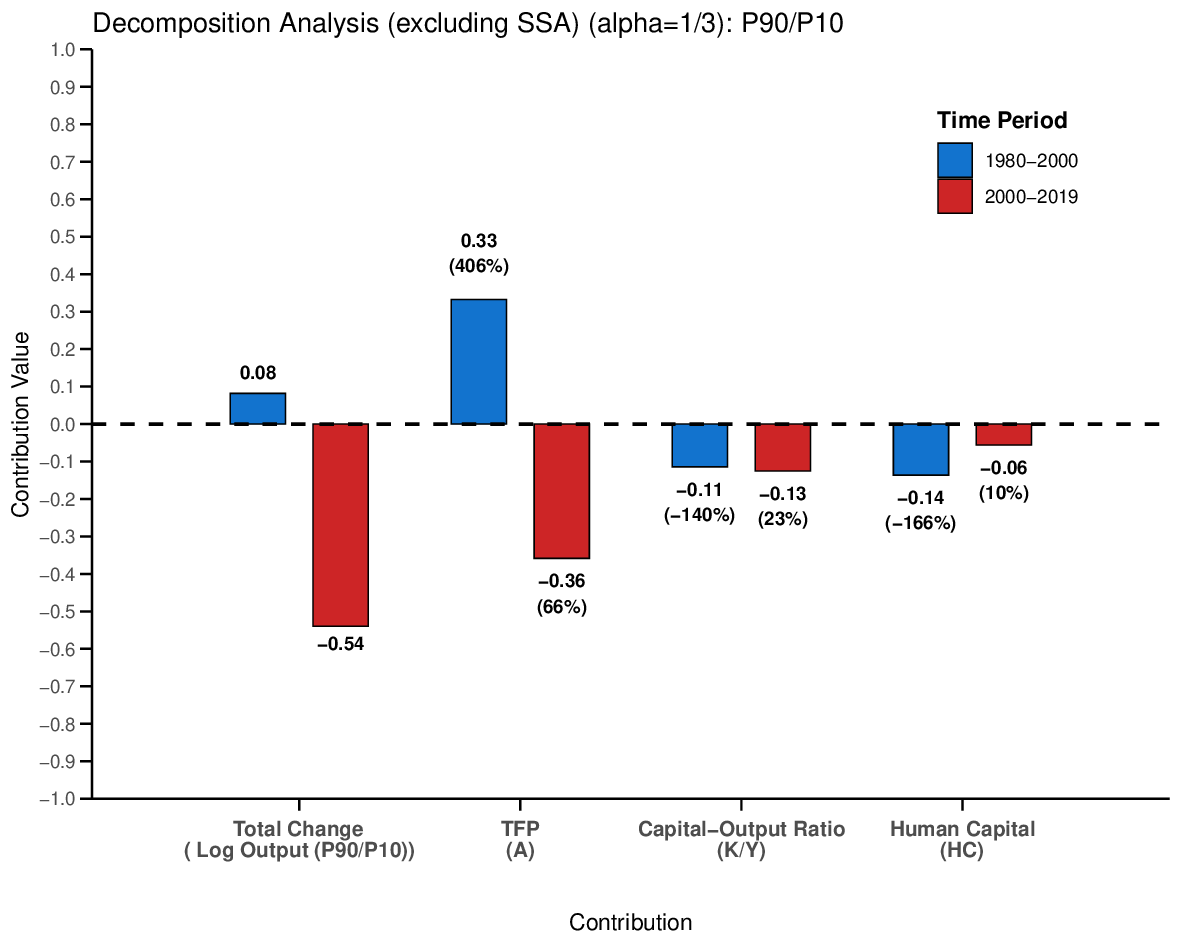}
\caption{Decomposition Exercise (excluding SSA, alpha=1/3): P90/P10}
\label{fig:decomposition_exercise_excluding_ssa_alpha_1_3_p90_p10}
\end{figure}

Figure \ref{fig_alpha_ptile} illustrates the capital share of income across countries at different percentiles of the income distribution for various years. For the full sample of countries included in this analysis, PWT shows an average capital share of income of 0.46, which significantly exceeds the commonly assumed value of 1/3 in the literature. Moreover, the capital share of income follows an inverse U-shaped pattern across the income distribution, with middle-income countries having the highest capital income share, followed by poor countries, and high-income countries having the lowest capital income share. Only the richest countries have capital income shares close to 1/3, suggesting that using a constant capital income share may not be appropriate for countries at other points in the income distribution. The decomposition approach in the current paper accommodates this heterogeneity, allowing the capital income share to vary along the income distribution. Additionally, over time, countries across the income distribution have experienced an increase in the capital share of income, which indicates a corresponding decline in labor share. The recent work by \cite{karabarbounis2014global} examines this global decline in labor share and attributes much of the trend to cheaper information and communication technology (ICT) capital. Because capital-income shares differ systematically across the income distribution, any growth-accounting exercise must incorporate this heterogeneity. Figure \ref{fig:decomposition_exercise_excluding_ssa_alpha_1_3_p90_p10} reports-for comparison—the results obtained when a constant capital share of one-third is imposed.

During the 1980-2000 period cross-country income divergence is still driven overwhelmingly by the widening gap in total factor productivity, with the convergence of physical and human capital acting in the opposite direction and therefore dampening the TFP effect. Because TFP remains the dominant force in the accounting, studies that concentrate on the pre-2000 era—such as \cite{klenow1997neoclassical}—are not materially affected by whether one allows the capital share of income to vary across countries or, instead, fixes it at the conventional one-third benchmark. However, for the post-2000 period, the choice of capital income share becomes important.  With a constant one-third share, the capital-output ratio accounts for only 23\% of the decline in the P90/P10 ratio, compared to 44\% when heterogeneity in capital income shares is incorporated. This difference implies that ignoring heterogeneity may lead to the erroneous conclusion that TFP remains the dominant factor in driving growth and convergence. Over the full period from 1980 to 2019, the constant one-third assumption results in capital convergence explaining 52\% of the overall convergence, as opposed to 89\% when capital income shares vary by income. Thus, even in the long run, assuming a constant capital share significantly underestimates the role of capital in driving income convergence between rich and poor economies.

The Appendix further presents the decomposition results for the P90/P50 and P50/P10 ratios under the constant one-third assumption, showing a quantitatively reduced role of capital in driving convergence compared to the case when capital income shares are allowed to vary. The results remains unchanged when looking at full sample of countries.

\section{Measurement of Inputs and Capital Share of Income}\label{sec_discussion}

The accounting exercise suggests that the convergence of inputs, particularly physical capital, has significantly driven overall convergence.  This section discusses how the measures of physical and human capital are constructed in the Penn World Tables (10.01) and evaluates the extent to which the accounting results depend on the assumptions underlying the construction of these variables.

\textbf{Physical Capital} \\
In this paper, the capital-output ratio is calculated as the ratio of real capital stock ($rnna$) to real GDP ($rgdpna$) at constant national prices from the Penn World Tables (PWT 10.01). The capital stock series in the PWT is constructed using the Perpetual Inventory Method (PIM). The stock of capital of type $a$ at time $t$ is computed as the accumulated type-specific investment ($I_{at}$) net of depreciation ($\delta_a$):
$$K_{at} = I_{at} + (1-\delta_a)K_{a, t-1}.$$ 
The aggregate capital stock is simply the sum of asset specific capital stocks, $K_{t} = \sum_a K_{at}$. The PIM requires an assumption about the initial capital stock ($K_0$), i.e., the capital stock in the first year when investment data becomes available. A common approach in the literature is to use the steady-state relation from the neoclassical growth model: 
$$K_0 = \frac{I_0}{\delta + g},$$ where $g$ is the steady state growth rate of investment. The PWT modifies this approach by using the steady-state assumption initially, then simulating the path of capital for all countries based on the PIM. It calculates cross-country median values of the capital-output ratio ($k_a$) and assumes all countries start with the same ratio. The initial capital stock is then recalculated as: $$K_{a0} = Y_0 k_{a0}.$$ This implies that differences in initial capital stocks are solely due to differences in initial output levels. However, this assumption is unlikely to affect the accounting results significantly. For countries in the main analysis, investment data is available since 1970 or earlier. Over time, the influence of the initial capital stock diminishes due to depreciation, especially for assets with high depreciation rates. Assuming a 5\% annual depreciation rate, on average only 5.7\% of capital in 2000 is due to underappreciated capital in 1970 for the countries included in the main analysis.\footnote{The capital in year $1970+k$ is given by $K_{1970+k} = \sum_{t=1}^{t=k}(1-\delta)^{t-1}I_t  + (1-\delta)^kK_{1970}$. The amount of capital in year $1970 +k$ due to undepreciated capital in year $1970$ is therefore given by $\frac{(1-\delta)^kK_{1970}}{K_{1970+k}}$. The amount of capital in year 1980, 1990, 2000, 2010, due to undepreciated capital in 1970 is 32.7\%, 13.6\%, 5.7\%, 2.5\% on average.}

Therefore, the decline in capital-output ratio differences as a key factor in reducing output differences post-2000 is not driven by assumptions about the initial capital stock. Instead, it is more likely driven by changes in investment patterns rather than initial capital stock differences.

\textbf{Human Capital} \\ 
The human capital measure in PWT10.01 is constructed using data on average years of schooling in the population aged 25 and older from \cite*{barro2013new}, \cite{cohen2007growth}, and assumed rates of return for primary, secondary, and tertiary education based on Mincer equation estimates from \cite*{psacharopoulos1994returns} and \cite{caselli2005accounting}. Specifically:
$$hc = exp\{\phi(s)\}$$ where 
\begin{equation*}
\phi(s) = 
\begin{cases}
0.134 \cdot s & \text{if  } s \leq 4;\\ 
0.134 \cdot 4 + 0.101 \cdot (s-4) & \text{if  } 4<s\leq 8; \\
0.134 \cdot 4 + 0.101 \cdot 4 + 0.068 \cdot (s-8) & \text{if  } s \geq 8
\end{cases}
\end{equation*}
This measure of human capital is imperfect because it does not account for the quality of schooling \citep*{schoellman2012education, hanushek2012better} or cross-country differences in returns to experience \citep*{lagakos2018life}. Since the returns to schooling are assumed to be constant across years, human capital convergence is solely driven by the convergence in years of schooling. If returns to schooling, the quality of education, or returns to experience are also converging across economies, then the contribution of human capital presented here is a lower bound. \\

\textbf{Capital Share of Income} \\ 
The capital share of income ($\alpha$) in PWT10.01 is derived using the following relationship: $$\alpha = 1 - labsh$$ where \textit{labsh} represents the share of labor income in GDP. One significant challenge in measuring the labor share is that the labor income of self-employed workers is not directly observable. \cite{gollin2002getting} discusses various methods for estimating the labor compensation of self-employed workers. 
Two common adjustments rely on mixed-income data from national accounts. Mixed income refers to the total income earned by self-employed workers, encompassing both labor and capital income. The first adjustment allocates all mixed income to labor, while the second adjustment allocates mixed income between labor and capital in the same proportion as the rest of the economy. In cases where mixed income data is unavailable, a third adjustment uses data on the number of employees and self-employed workers, assuming that the self-employed earn the same average wage as employees. PWT10.01 introduces a fourth adjustment that adds the value-added from agriculture to labor compensation of employees. Whenever mixed income data is available, PWT10.01 applies the second adjustment. In the absence of mixed-income data, PWT10.01 uses the minimum of the third and fourth adjustments.

\section{Conclusion}\label{sec_conclusion} 
This study revisited the dynamics of cross-country income convergence using the most recent data from the Penn World Tables (PWT 10.01).\footnote{ Roughly 70\% of cross-country empirical work relies on the Penn World Tables \citep{johnson2020remains}.} Contrary to the traditional view that emphasizes non-convergence and the limited role of factor inputs, the analysis reveals a significant shift in global income patterns since 2000. There has been unconditional convergence, particularly outside Sub-Saharan Africa. More importantly, accumulation of inputs, particularly physical capital, has significantly driven income convergence outside Sub-Saharan Africa.\\ 
This suggests that the factors determining global income dynamics have changed dramatically, necessitating a reevaluation of existing growth theories and models. The results open several avenues for future research. What factors have contributed to the reduction in input dispersion? What underlies the persistent stagnation of Sub-Saharan economies despite their market-oriented reforms? 
The findings also have significant policy implications. Since input accumulation, rather than TFP growth, primarily drives convergence, policies aimed at removing barriers to the accumulation of physical and human capital in poorer economies may yield more substantial growth dividends than those focused solely on TFP enhancements.

\bmhead{Acknowledgements}

I am grateful to Anmol Bhandari, Alessandra Fogli, Hannes Malmberg, Todd Schoellman, and  Kjetil Storesletten for their advice and support. I would also like to thank Honey Batra, Francisco Bullano, G V A Dharanan, Jason Hall, Sang Min Lee, Ruyan Liao, Martin Garcia Vazquez, and  Teerat Wongrattanapiboon for useful comments.

\begin{appendices}

    \section{Mathematical Derivations}\label{app_maths}
\subsection{Generalized Decomposition}\label{derivation_gen_deom}

Let $p$ denote the percentile index, and consider the per‐worker output function
\begin{equation}\label{eq:basic}
  y(p)
  =F\bigl(A(p),\,k(p),\,h(p)\bigr)
  =F\bigl(A(p),\,(k(p)/y(p))\,y(p),\,h(p)\bigr),
\end{equation}
where $\kappa(p)=k(p)/y(p)$.  Total differentiation of \eqref{eq:basic} with respect to $p$ gives
\begin{align}
  \frac{dy}{dp}
  &=F_{A}\bigl(A,\kappa y,h\bigr)\,\frac{dA}{dp}
    +F_{k}\bigl(A,\kappa y,h\bigr)\,\frac{d(\kappa y)}{dp}
    +F_{h}\bigl(A,\kappa y,h\bigr)\,\frac{dh}{dp}, \label{eq:dy_dp1}\\
  \frac{d(\kappa y)}{dp}
  &=\kappa\,\frac{dy}{dp}+y\,\frac{d\kappa}{dp}. \label{eq:dykappa_dp}
\end{align}
Rearranging yields
\begin{equation}\label{eq:dy_dp2}
  \bigl[1 - F_{k}(A,\kappa y,h)\,\kappa\bigr]\,\frac{dy}{dp}
  =F_{A}(A,\kappa y,h)\,\frac{dA}{dp}
   +F_{k}(A,\kappa y,h)\,y\,\frac{d\kappa}{dp}
   +F_{h}(A,\kappa y,h)\,\frac{dh}{dp}.
\end{equation}
Dividing by $y$ gives the differential form
\begin{equation}\label{eq:dlny_dp}
  \frac{d\ln y}{dp}
  =\underbrace{\frac{F_{A}}{y\,\bigl(1 - F_{k}\,\kappa\bigr)}\,\frac{dA}{dp}}_{\mathcal{E}_{A}(p)}
  +\underbrace{\frac{F_{k}\,\kappa}{1 - F_{k}\,\kappa}\,\frac{d\ln\kappa}{dp}}_{\mathcal{E}_{k/y}(p)}
  +\underbrace{\frac{F_{h}\,h}{y\,\bigl(1 - F_{k}\,\kappa\bigr)}\,\frac{d\ln h}{dp}}_{\mathcal{E}_{h}(p)}.
\end{equation}
Integration from $p_{0}$ to $p_{n}$ yields the generalized decomposition
\[
  \ln y(p_{n}) - \ln y(p_{0})
  =\int_{p_{0}}^{p_{n}}[\mathcal{E}_{A}(p)
    +\mathcal{E}_{k/y}(p)
    +\mathcal{E}_{h}(p)]\,dp.
\]

For a CRS technology in $(k,h)$, Euler’s theorem implies
\begin{equation}\label{eq:CRS}
  F_{k}(A,k,h)\,k + F_{h}(A,k,h)\,h = y,
  \quad
  \alpha(p) = \frac{F_{k}(A,k,h)\,k}{y}.
\end{equation}
Hence
\[
  F_{k}(A,k,h)\,\kappa = \alpha(p),
  \quad
  \frac{F_{h}(A,k,h)\,h}{y} = 1-\alpha(p).
\]
Using these identities, the three elasticities in \eqref{eq:dlny_dp} become
\begin{align*}
  \mathcal{E}_{A}(p)
  &= \frac{F_{A}}{y\,(1-\alpha(p))}\,\frac{dA}{dp}
   = \frac{1}{1-\alpha(p)}\,\frac{d\ln A(p)}{dp},\\
  \mathcal{E}_{k/y}(p)
  &= \frac{\alpha(p)}{1-\alpha(p)}\,\frac{d\ln(k/y)(p)}{dp},\\
  \mathcal{E}_{h}(p)
  &= \frac{1-\alpha(p)}{1-\alpha(p)}\,\frac{d\ln h(p)}{dp}
   = \frac{d\ln h(p)}{dp}.
\end{align*}
Substituting back into the integrated form yields the final result
\begin{equation}\label{eqn_decomp}
  \ln y(p_{n}) - \ln y(p_{0})
  =\int_{p_{0}}^{p_{n}}\Bigl[\frac{1}{1-\alpha(p)}\,\frac{d\ln A(p)}{dp}
    +\frac{\alpha(p)}{1-\alpha(p)}\,\frac{d\ln(k/y)(p)}{dp}
    +\frac{d\ln h(p)}{dp}\Bigr]\,dp.
\end{equation}

    The contribution of $A$ can be calculated as a residual. 
    The contribution of each of the inputs can then be evaluated as follows:
    \begin{align*}
    Contri_{\log(k/y)}  & = \int_{p_0}^{p_n}\frac{\alpha(p)}{1-\alpha(p)} \frac{d \log (k/y)(p)}{dp} dp \\
    & \approx  \sum_{k=0}^{n-1} \frac{\frac{\alpha(p_{k+1})}{1-\alpha(p_{k+1})} + \frac{\alpha(p_{k})}{1-\alpha(p_{k})}}{2} [\log(k/y)(p_{k+1}) - \log(k/y)(p_{k})] \\
    Contri_{\log(h)}  & = \int_{p_0}^{p_n} \frac{d \log (h)}{dp} dp \\
    & = \log(h(p_n)) - \log(h(p_0)) \\
    Contri_{\log(A)} & =  \int_{p_0}^{p_n} [\frac{d \log(y)}{dp} - \frac{\alpha(p)}{1-\alpha(p)} \frac{d \log (k/y)(p)}{dp} -  \frac{d \log (h)}{dp} ]dp
    \end{align*}
    
    Note that if the elasticity parameter is constant along income distribution ( $\alpha(p)= \alpha$), the generalized decomposition reduces to the usual Hall-Jones decomposition. 
    \begin{align*}
    \ln(y(p_n)) - \ln(y(p_0)) & = \frac{1}{1-\alpha}  \ln(\frac{A(p_n)}{A(p_0)}) + \frac{\alpha}{1-\alpha} \ln(\frac{k/y(p_n)}{k/y(p_0)}) + \ln(\frac{h(p_n)}{h(p_0)}).
    \end{align*}

    \newpage
    \section{Robustness and Additional Exercises}\label{app_robustness}

    \subsection{Robustness to different measures of GDP}
    \begin{table}[htbp]
      \centering
      \caption{Beta–Convergence Coefficients Across GDP Measures}
      \label{tab:beta_comparision}
      \begin{tabular}{@{} ll rr @{}}
        \toprule
        & & \multicolumn{2}{c}{GDP Measure} \\
        \cmidrule(l){3-4}
        & & RGDPO & RGDPE \\
        \midrule
        \multicolumn{4}{l}{\textbf{Panel A: All Countries}}\\
        Current sample & $\beta$ & \textbf{–0.0062} & –0.0052 \\
                       & (SE)    & (0.0017)        & (0.0016) \\
                       & $N$     & 155             & 155      \\
        PSS sample     & $\beta$ & –0.0052         & \textbf{–0.0043} \\
                       & (SE)    & (0.0016)        & (0.0016) \\
                       & $N$     & 124             & 124      \\
        \addlinespace
        \multicolumn{4}{l}{\textbf{Panel B: Outside SSA}}\\
        Current sample & $\beta$ & \textbf{–0.0150} & –0.0128 \\
                       & (SE)    & (0.0023)        & (0.0022) \\
                       & $N$     & 115             & 115      \\
        PSS sample     & $\beta$ & –0.0155         & \textbf{–0.0135} \\
                       & (SE)    & (0.0021)        & (0.0021) \\
                       & $N$     & 90              & 90       \\
        \bottomrule
      \end{tabular}
    
      \vspace{0.5ex}
      {\footnotesize
      \textbf{Notes:} $\beta$ coefficients are estimated by nonlinear least squares as in \eqref{eqn_reg_beta}, with robust standard errors in parentheses.  RGDPO and RGDPE denote output-side and expenditure-side real GDP per capita at chained PPPs (million 2017 US\$) from PWT 10.01. \emph{PSS sample} refers to the sample of country used in \cite{patel2021new}.
      }
    \end{table}
    
    The Penn World Table (PWT) provides two conceptually distinct real-GDP series: an output-side measure (RGDPO) and an expenditure-side measure (RGDPE).  \citet{feenstra2015next} recommend employing RGDPE when the objective is to compare welfare across countries and over time, whereas RGDPO is more appropriate for analysing productive capacity.  Table \ref{tab:beta_comparision} tests whether the choice between these two series affects estimates of $\beta$-convergence over 2000-2019.  The table reports coefficients for both the  current sample and the country set used by \citet{patel2021new} (PSS sample), and it distinguishes between the full set of countries and the subset that excludes Sub-Saharan Africa (SSA).
    
    Across all countries the estimated convergence rate lies between \(-0.0062\) (output-side GDP, current sample) and \(-0.0043\) (expenditure-side GDP, PSS sample); all coefficients are negative and statistically significant at the one-percent level.  Restricting attention to economies outside SSA roughly triples the speed of convergence, with point estimates ranging from \(-0.0155\) to \(-0.0128\).   It can thus be concluded that $\beta$‑income convergence remains robust to the choice between RGDPO and RGDPE.

    \section{Additional Figures}\label{app_figs}
    
    \begin{figure}[htbp]
      \centering
      \includegraphics[width=\linewidth]{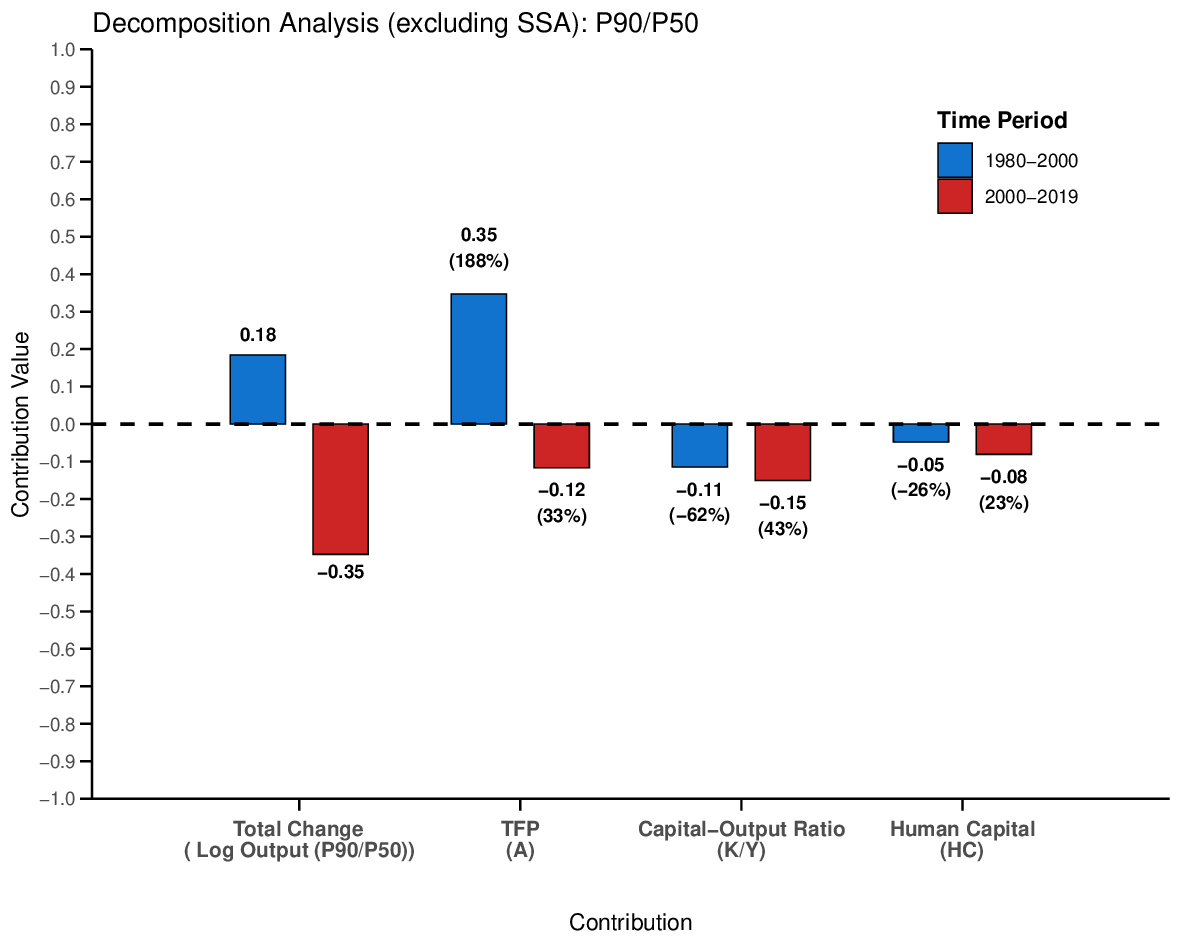}
      \caption{Decomposition Analysis (excluding SSA): P90/P50}
      \label{fig:decomposition_analysis_excluding_ssa_p90_p50}
    
      \vspace{0.5ex}
      {\footnotesize
      \textbf{Notes:} The figure decomposes the change in the log P90/P50 output ratio into the contributions of total factor productivity, the capital-output ratio, and human capital for the periods 1980-2000 and 2000-2019 for countries outside Sub-Saharan Africa. Inputs (capital-output ratio and human capital) account for 66\% of the convergence in the log P90/P50 ratio between 2000-2019 and for the entire convergence over 1980-2019.
      }
    \end{figure}

    \begin{figure}[htbp]
      \centering
      \includegraphics[width=\linewidth]{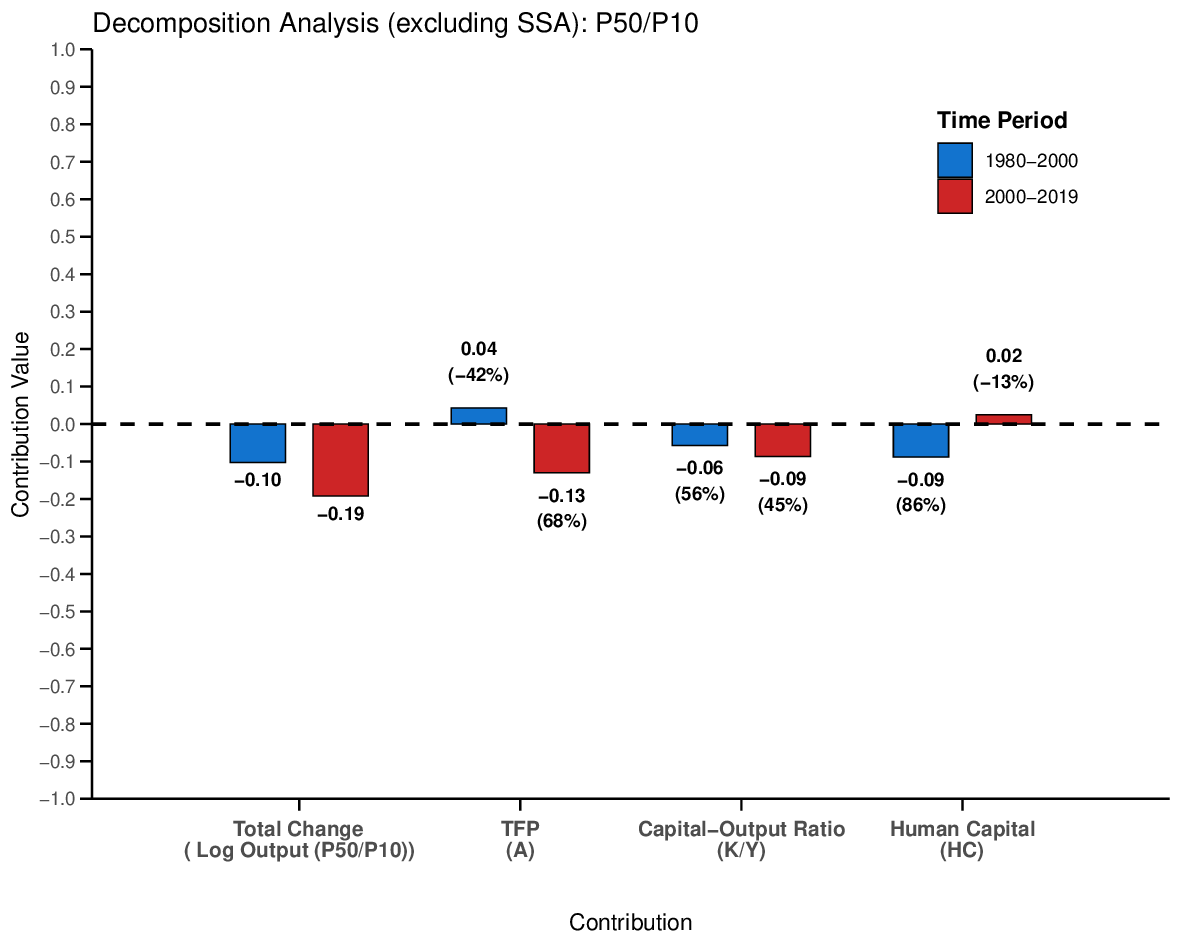}
      \caption{Decomposition Analysis (excluding SSA): P50/P10}
      \label{fig:decomposition_analysis_excluding_ssa_p50_p10}
    
      \vspace{0.5ex}
      {\footnotesize
      \textbf{Notes:} The figure decomposes the change in the log P50/P10 output ratio into the contributions of total factor productivity, the capital-output ratio, and human capital for the periods 1980-2000 and 2000-2019 for countries outside Sub-Saharan Africa.  Inputs (capital-output ratio and human capital) account for 32\% of the convergence in the log P50/P10 ratio between 2000-2019 and for 70\% of the convergence over 1980-2019.
      }
    \end{figure}

    \begin{figure}[htbp]
      \centering
      \includegraphics[width=\linewidth]{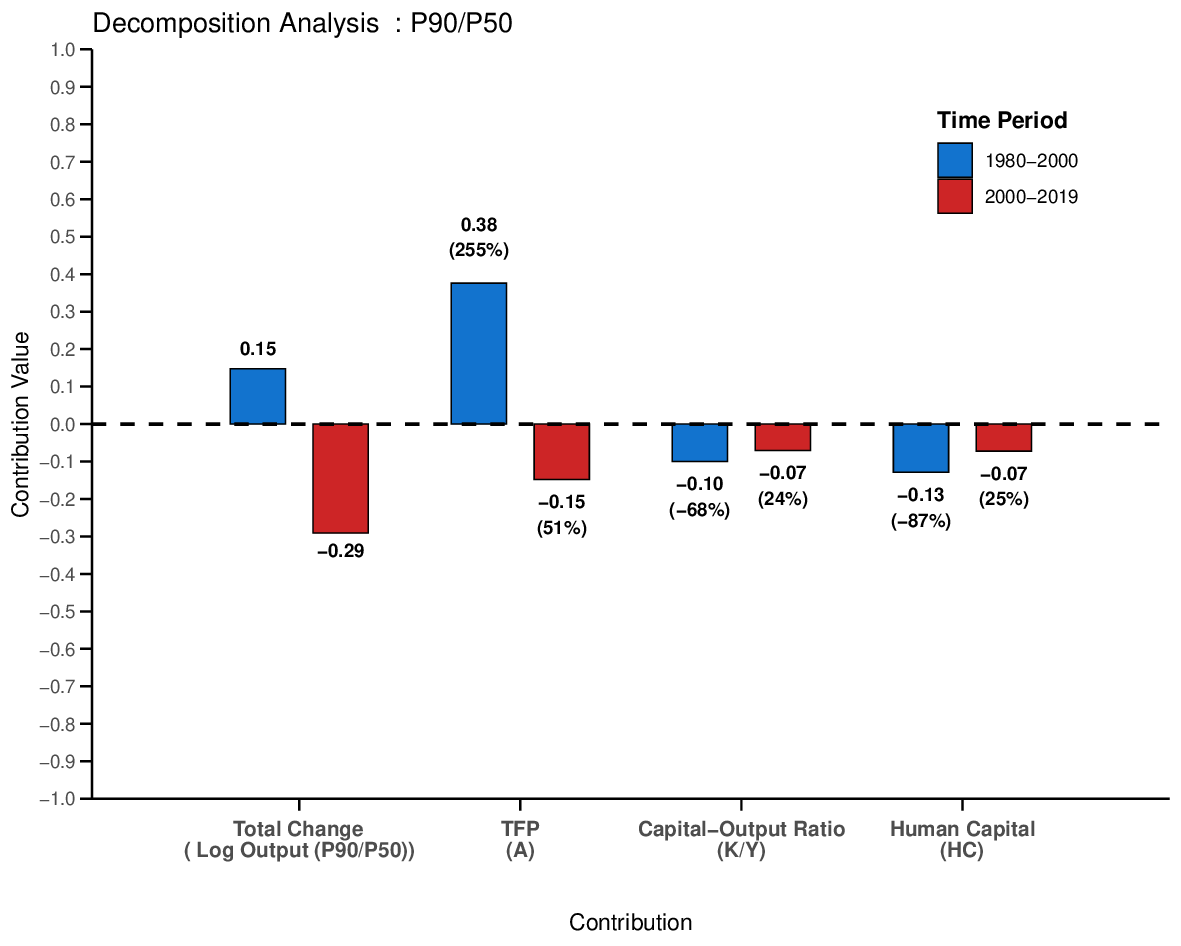}
      \caption{Decomposition Analysis (full sample): P90/P50}
      \label{fig:decomposition_analysis_full_sample_p90_p50}
    
      \vspace{0.5ex}
      {\footnotesize
      \textbf{Notes:} The figure decomposes the change in the log P90/P50 output ratio into the contributions of total factor productivity, the capital-output ratio, and human capital for the periods 1980-2000 and 2000-2019 using the full sample of countries.  Inputs (capital-output ratio and human capital) account for 49\% of the convergence in the log P90/P50 ratio between 2000-2019 and for the entire convergence over 1980-2019.
      }
    \end{figure}
    
    \begin{figure}[htbp]
      \centering
      \includegraphics[width=\linewidth]{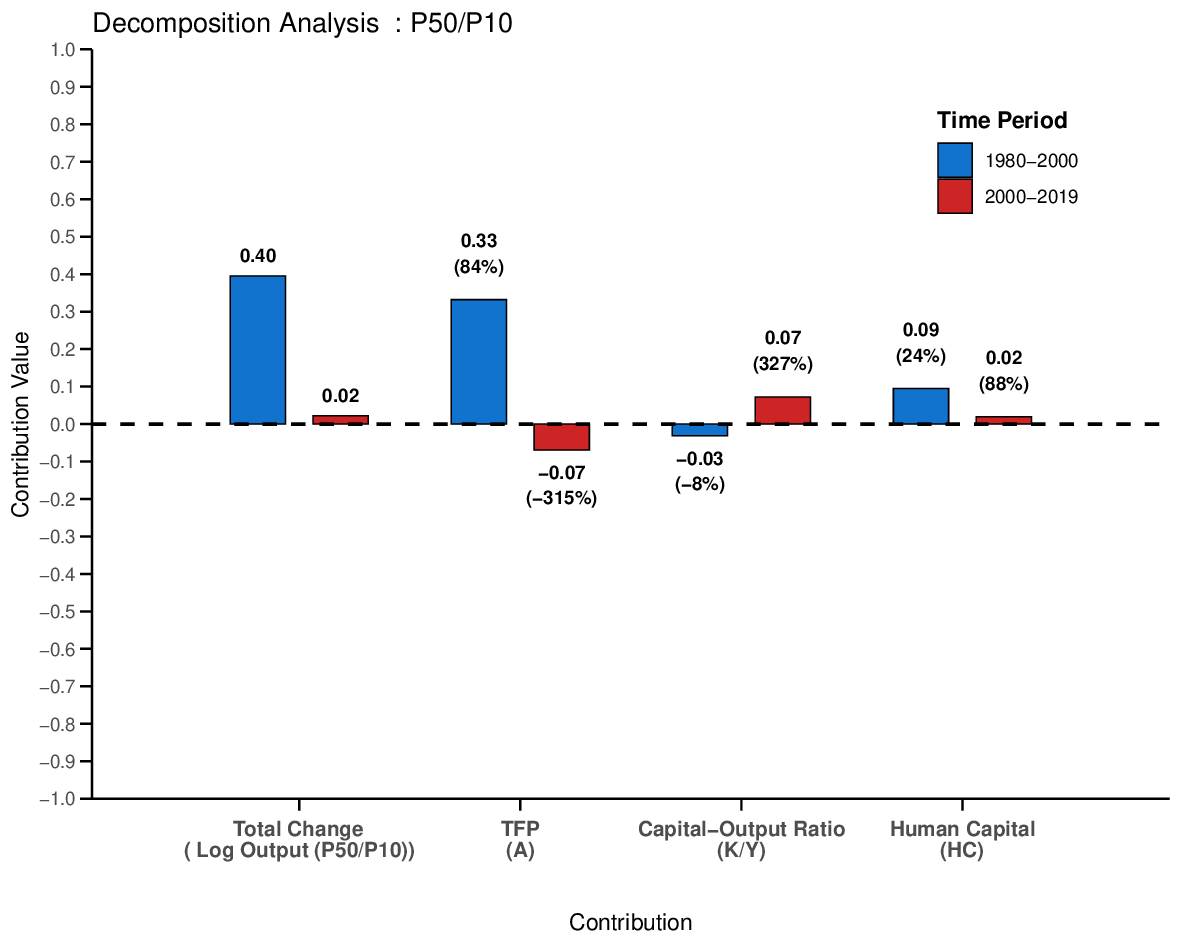}
      \caption{Decomposition Analysis (full sample): P50/P10}
      \label{fig:decomposition_analysis_full_sample_p50_p10}
    
      \vspace{0.5ex}
      {\footnotesize
      \textbf{Notes:} The figure decomposes the change in the log P50/P10 output ratio into the contributions of total factor productivity, the capital-output ratio, and human capital for the periods 1980-2000 and 2000-2019 using the full sample of countries.  During 2000-2019 inputs (capital-output ratio and human capital) diverged, whereas TFP exhibited a slight convergence, yielding a net divergence in the P50/P10 ratio.  Over the entire horizon 1980-2019 TFP divergence explains 63 \% of the overall income divergence, while input divergence accounts for a further 27 \%.
      }
    \end{figure}

    
    \begin{figure}[htbp]
      \centering
      \includegraphics[width=\linewidth]{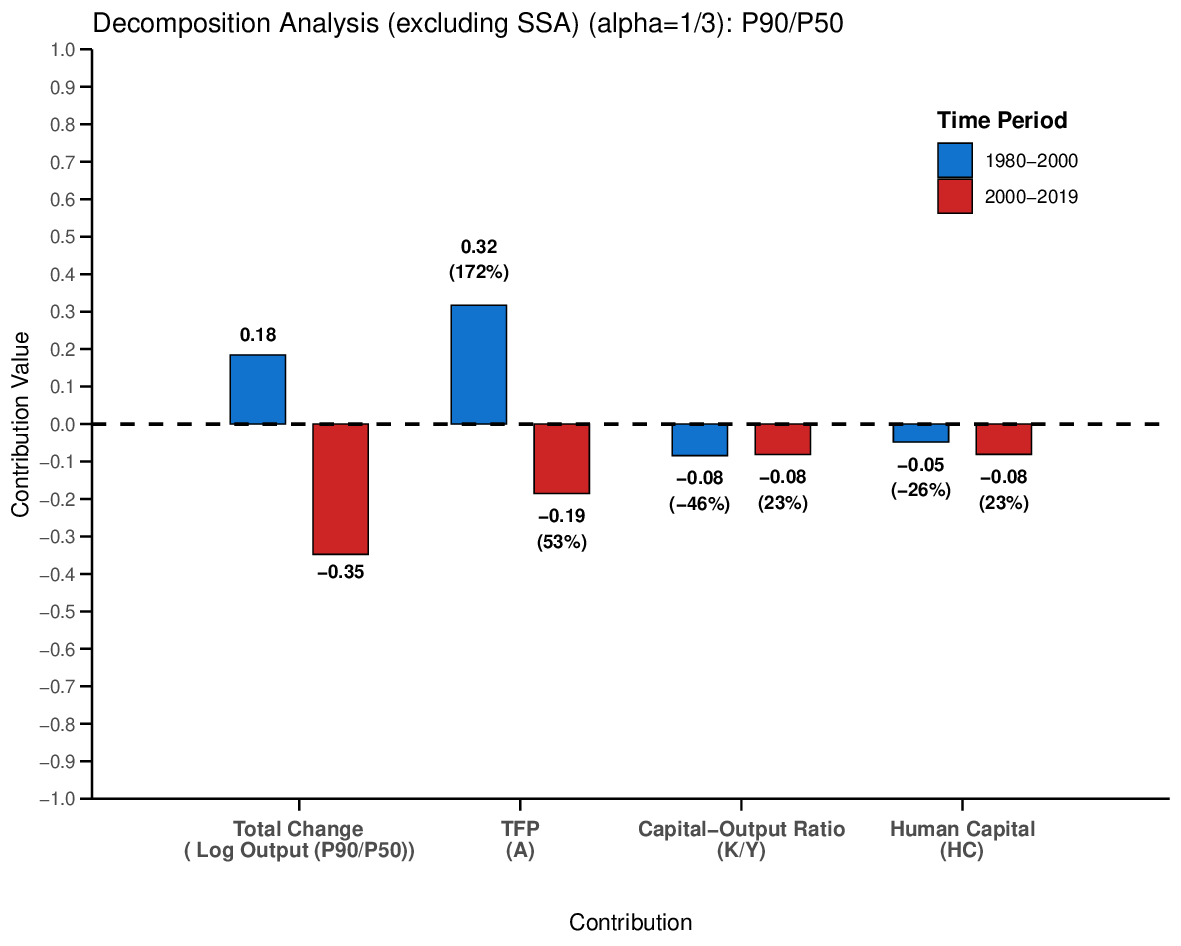}
      \caption{Decomposition Analysis (excluding SSA, $\alpha = 1/3$): P90/P50}
      \label{fig:decomposition_analysis_excluding_ssa_alpha_1_3_p90_p50}
    
      \vspace{0.5ex}
      {\footnotesize
      \textbf{Notes:} The figure decomposes the change in the log P90/P50 output ratio into the contributions of total factor productivity, the capital-output ratio, and human capital for 1980--2000 and 2000--2019, assuming a constant capital-income share of one-third.  Inputs (capital--output ratio and human capital) account for 47\% of the convergence in 2000--2019 and for the entire convergence over 1980--2000.
      }
    \end{figure}
    
    \begin{figure}[htbp]
      \centering
      \includegraphics[width=\linewidth]{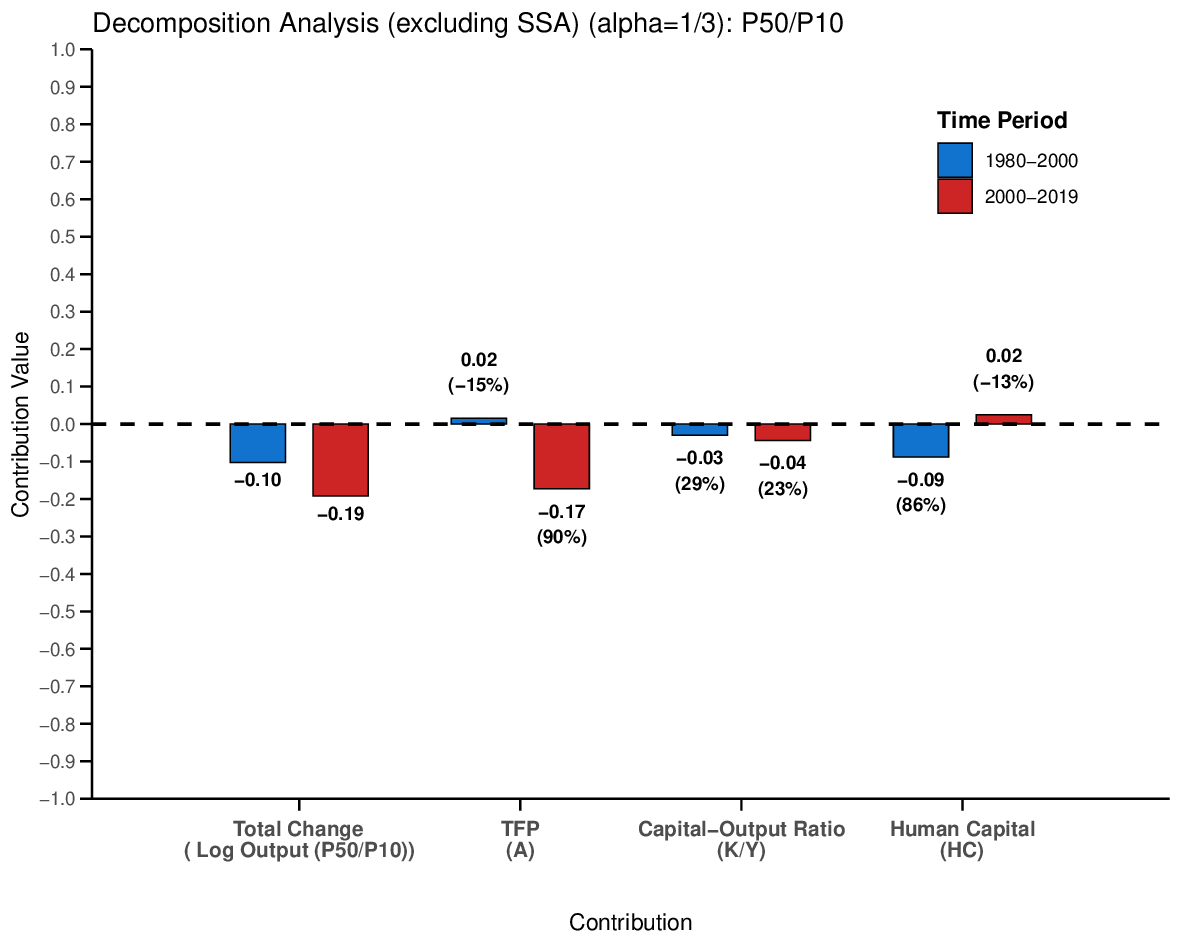}
      \caption{Decomposition Analysis (excluding SSA, $\alpha = 1/3$): P50/P10}
      \label{fig:decomposition_analysis_excluding_ssa_alpha_1_3_p50_p10}
    
      \vspace{0.5ex}
      {\footnotesize
      \textbf{Notes:} The figure decomposes the change in the log P50/P10 output ratio into the contributions of total factor productivity, the capital--output ratio, and human capital for 1980--2000 and 2000--2019, assuming a constant capital-income share of one-third.  Inputs account for 10\% of the convergence in 2000--2019 and for 37\% of the convergence over 1980--2000.
      }
    \end{figure}
    
    
    \begin{figure}[htbp]
      \centering
      \includegraphics[width=\linewidth]{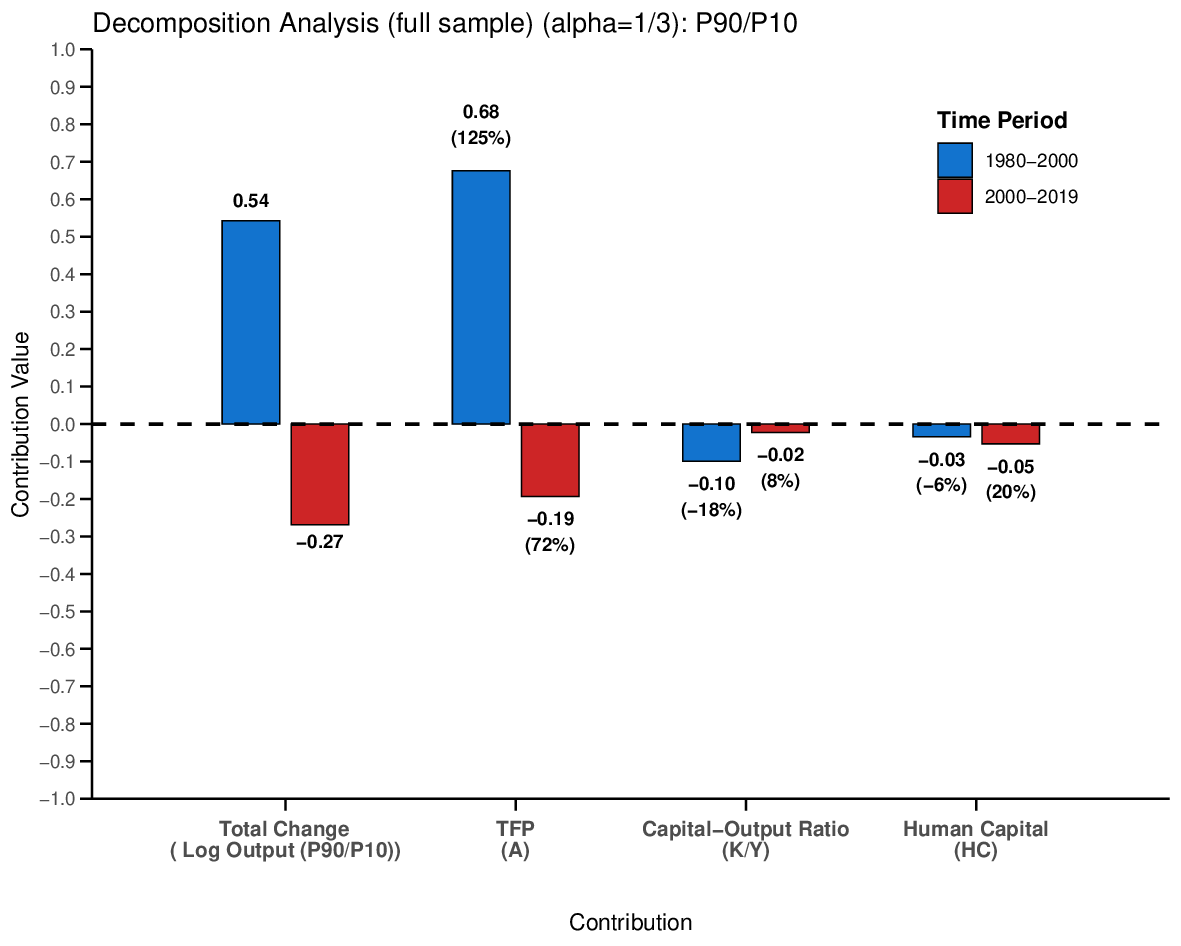}
      \caption{Decomposition Analysis (full sample, $\alpha = 1/3$): P90/P10}
      \label{fig:decomposition_analysis_full_sample_alpha_1_3_p90_p10}
    
      \vspace{0.5ex}
      {\footnotesize
      \textbf{Notes:} The figure decomposes the change in the log P90/P10 output ratio into the contributions of total factor productivity, the capital--output ratio, and human capital for 1980--2000 and 2000--2019, assuming a constant capital-income share of one-third.  Capital convergence accounts for 28\% of the convergence in 2000--2019.  Over the full 1980--2019 horizon the ratio diverges overall, driven by large TFP divergence in 1980--2000; capital convergence mitigates but does not overturn this early divergence.
      }
    \end{figure}
    
    \begin{figure}[htbp]
      \centering
      \includegraphics[width=\linewidth]{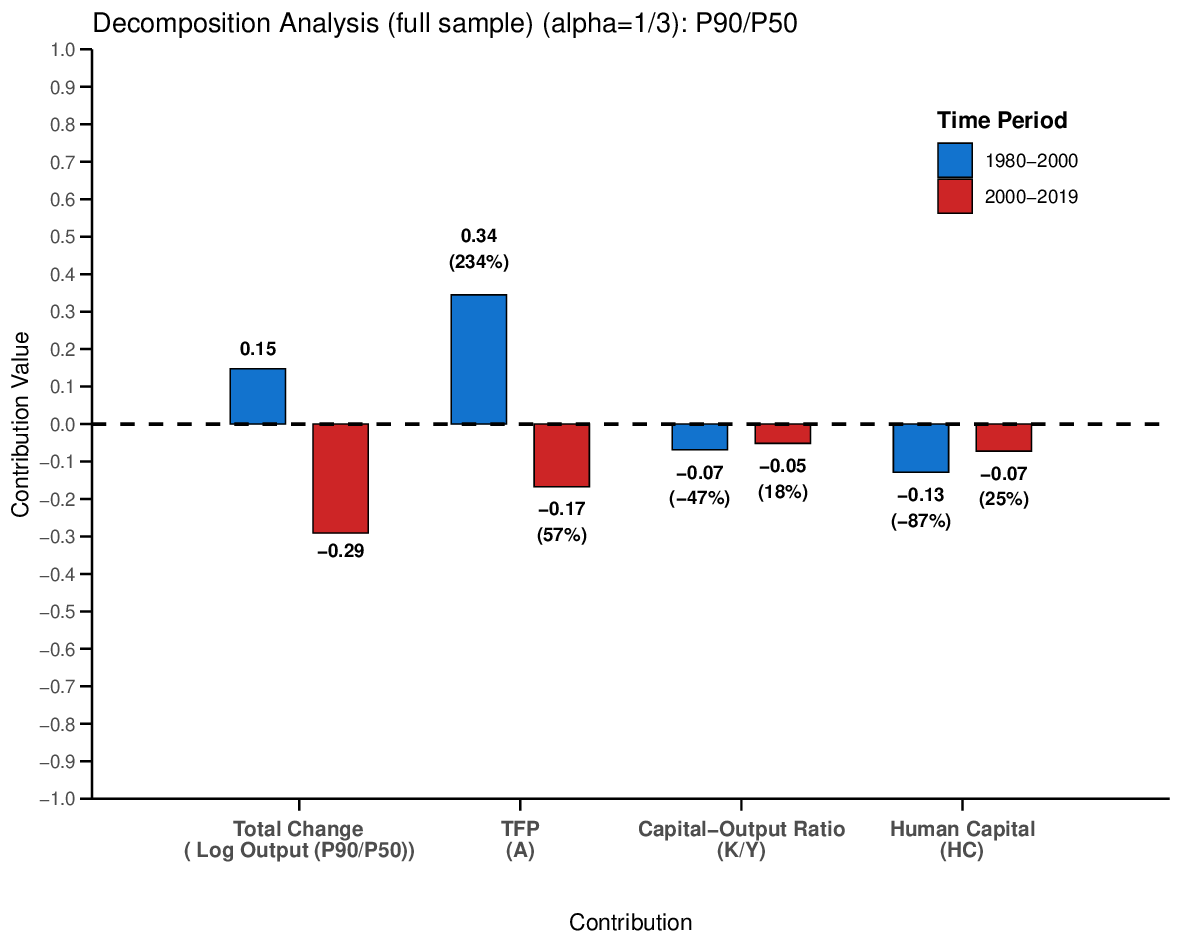}
      \caption{Decomposition Analysis (full sample, $\alpha = 1/3$): P90/P50}
      \label{fig:decomposition_analysis_full_sample_alpha_1_3_p90_p50}
    
      \vspace{0.5ex}
      {\footnotesize
      \textbf{Notes:} The figure decomposes the change in the log P90/P50 output ratio into the contributions of total factor productivity, the capital--output ratio, and human capital for 1980--2000 and 2000--2019, assuming a constant capital-income share of one-third.  Inputs account for 42.6\% of the convergence in 2000--2019 and for the entire convergence over 1980--2019.
      }
    \end{figure}
    
    \begin{figure}[htbp]
      \centering
      \includegraphics[width=\linewidth]{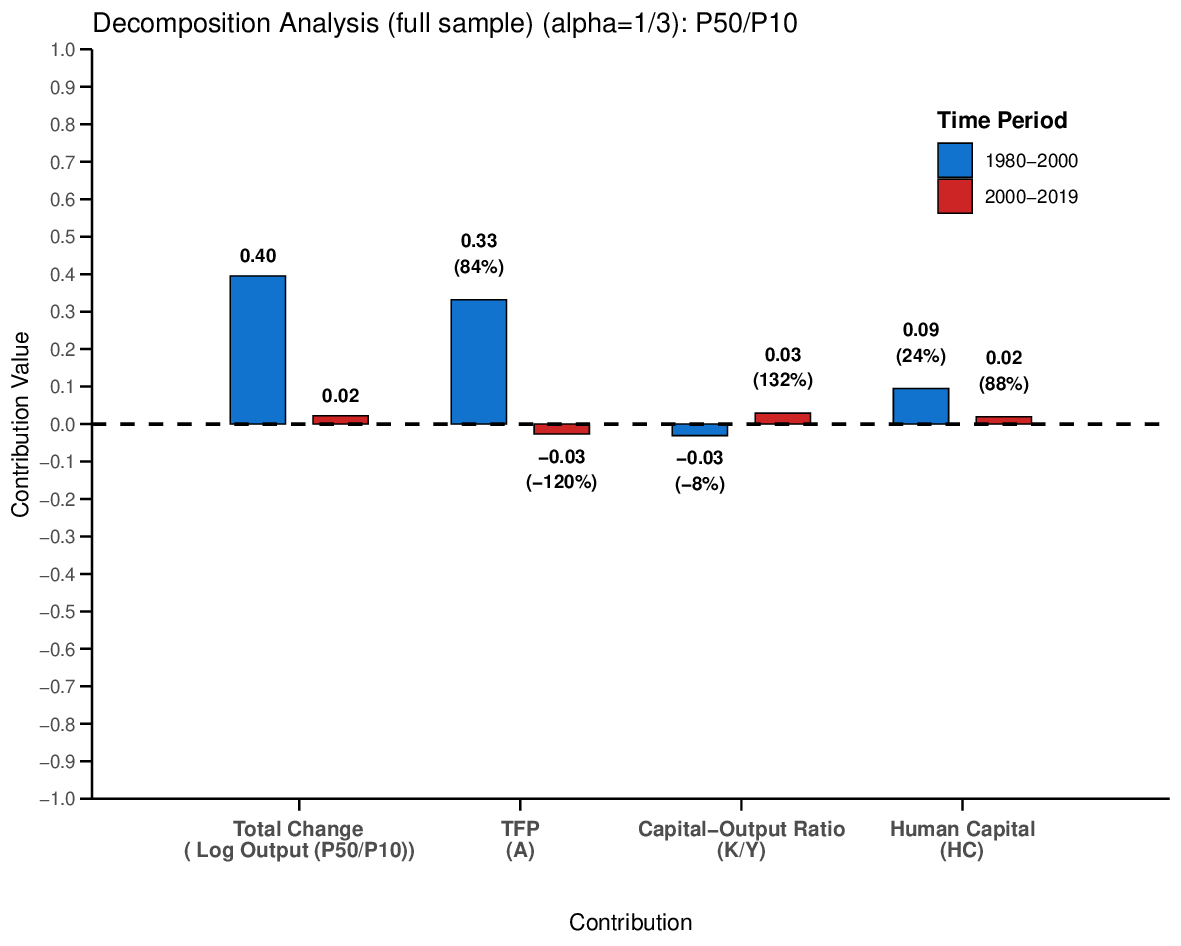}
      \caption{Decomposition Analysis (full sample, $\alpha = 1/3$): P50/P10}
      \label{fig:decomposition_analysis_full_sample_alpha_1_3_p50_p10}
    
      \vspace{0.5ex}
      {\footnotesize
      \textbf{Notes:} The figure decomposes the change in the log P50/P10 output ratio into the contributions of total factor productivity, the capital--output ratio, and human capital for 1980--2000 and 2000--2019, assuming a constant capital-income share of one-third.  In 2000--2019 overall output diverges despite TFP convergence because of divergence in capital inputs.  Over 1980--2019 both TFP and capital diverge, reinforcing the long-run divergence in the P50/P10 ratio.
      }
    \end{figure}

    \clearpage
    \subsection{Variance Decomposition}\label{app_var_decom}
    This section complements the main accounting analysis by performing the accounting based on variance as a measure of income dispersion, which is more commonly used in the literature. However, I prefer the percentile decomposition over variance decomposition for several reasons. First, percentile ratios allow for continuous variation in parameters across time and countries. Second, percentile ratios are less sensitive to outliers compared to variance measures. Furthermore, variance decomposition includes a term for the covariance between total factor productivity (TFP) and inputs, which can be difficult to interpret. For this exercise, I use a Cobb-Douglas production function, and a capital share parameter ($\alpha$) of 0.46, which is the observed mean in PWT10.01.
    
    Suppose we have a Cobb-Douglas production function. 
    $$Y = \tilde{A} K^{\alpha}(hL)^{1-\alpha}$$ 
    This can be rewritten as 
    $$\frac{Y}{L} = A \left(\frac{K}{Y}\right)^{\frac{\alpha}{1-\alpha}} h,$$ 
    or 
    $$y = A\, y^{kh},$$  
    where $y = \frac{Y}{L}$ and $y^{kh} = \left(\frac{K}{Y}\right)^{\frac{\alpha}{1-\alpha}} h.$ Consequently,
    $$Var(\ln(y)) = Var(\ln(A)) + Var(\ln(y^{kh})) + 2\,Cov(\ln(A), \ln(y^{kh})).$$

    \begin{figure}[htbp]
      \centering
      \includegraphics[width=\linewidth]{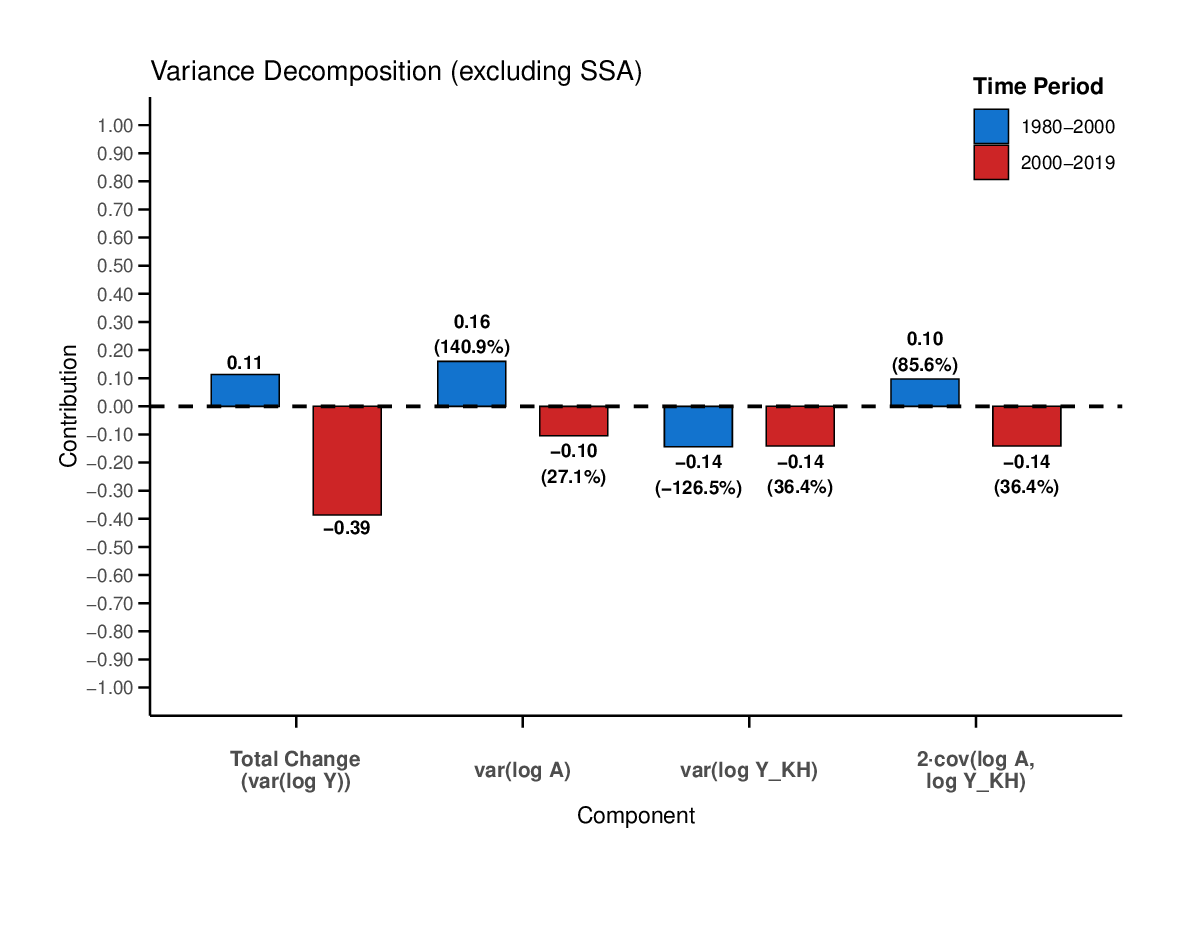}
      \caption{Variance Decomposition Outside Sub-Saharan Africa}
      \label{fig:variance_decomposition_excluding_ssa}
    
      \vspace{0.5ex}
      {\footnotesize
      \textbf{Notes:} The figure decomposes the change in the variance of log output per worker into the contributions of total factor productivity (TFP), the combined variance of physical- and human-capital inputs, and the covariance between TFP and inputs for the intervals 1980-2000 and 2000-2019 for countries outside Sub-Saharan Africa.  
      }
    \end{figure}
    
    \begin{figure}[htbp]
      \centering
      \includegraphics[width=\linewidth]{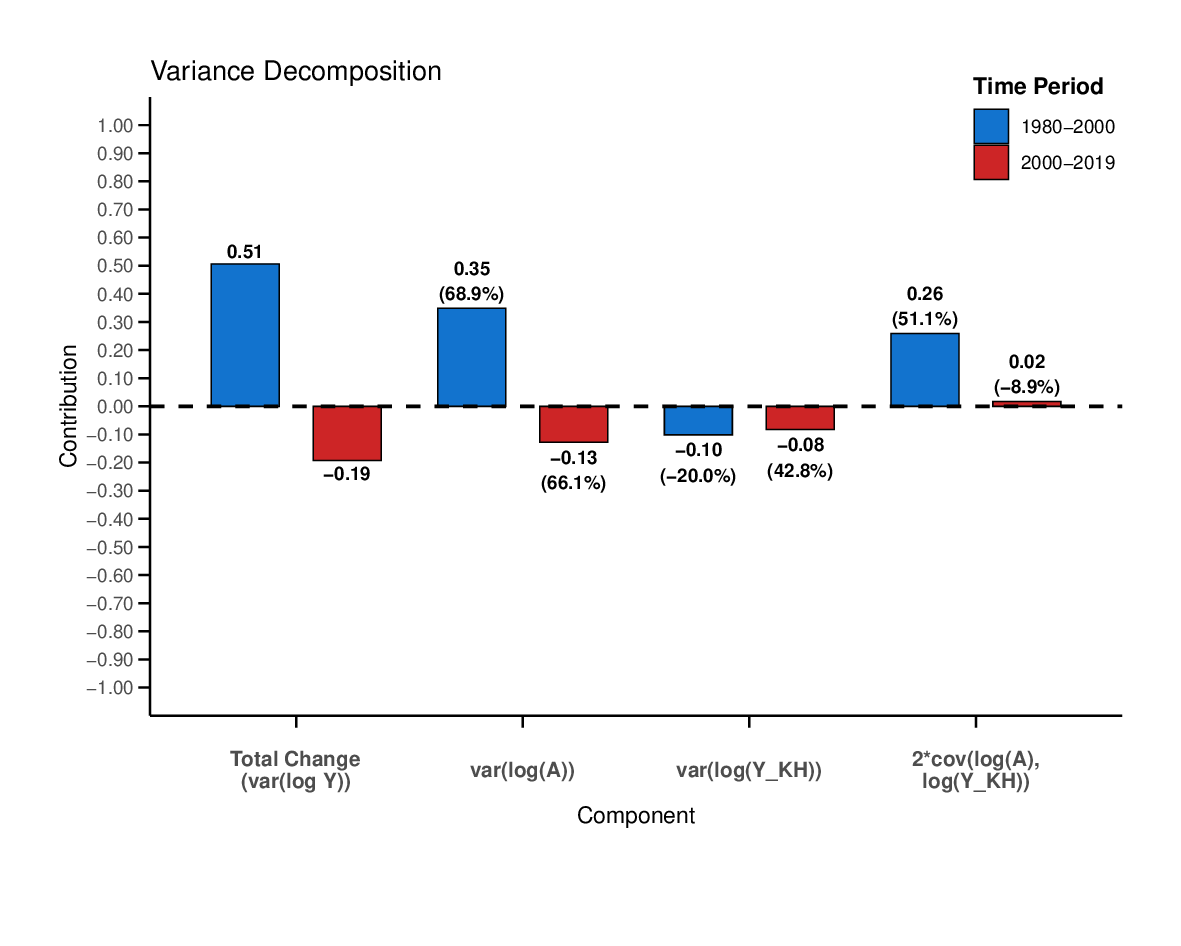}
      \caption{Variance Decomposition, Full Country Sample}
      \label{fig:variance_decomposition_full_sample}
    
      \vspace{0.5ex}
      {\footnotesize
      \textbf{Notes:} The figure reports the same variance decomposition for the full set of countries. 
      }
    \end{figure}

    Figures \ref{fig:variance_decomposition_excluding_ssa} and \ref{fig:variance_decomposition_full_sample} trace how the dispersion of income per worker has evolved by separating the variance of log output into three components: the variance of total factor productivity, the variance of combined physical- and human-capital inputs, and their covariance.
    
    For countries outside Sub-Saharan Africa, the variance of log output per worker rose by 11 percent between 1980 and 2000. TFP divergence contributed 141 percent of this increase, and the covariance between TFP and inputs added a further 86 percent, whereas convergence in physical and human capital offset 126 percent of the potential rise. In the subsequent period, 2000-2019, the variance fell by 39 percent: convergence in inputs accounted for 36 percent of that reduction, the fall in the covariance term accounted for another 36 percent, and TFP convergence  accounted for the remainder. Taken together, these forces produced a net 27-percent decline in the variance over 1980-2019, with input convergence contributing 104 percent of the total reduction.
    
    For the full global sample, the variance of log output increased by 51 percent between 1980 and 2000, with TFP divergence responsible for 69 percent of the rise and the covariance term for 51 percent, while input convergence offset 20 percent. During 2000-2019 the variance contracted by 19 percent: TFP convergence contributed 66 percent of the decline, input convergence 43 percent, and a further expansion of the covariance term offset 9 percent. Over the entire horizon 1980-2019 the variance remained 31 percent above its initial level; TFP divergence accounted for 71 percent of this long-run increase and the covariance term for 88 percent, whereas input convergence reduced the total by 59 percent.

    \clearpage 
    
    \begin{figure}[htbp]
      \centering
      \includegraphics[width=\linewidth]{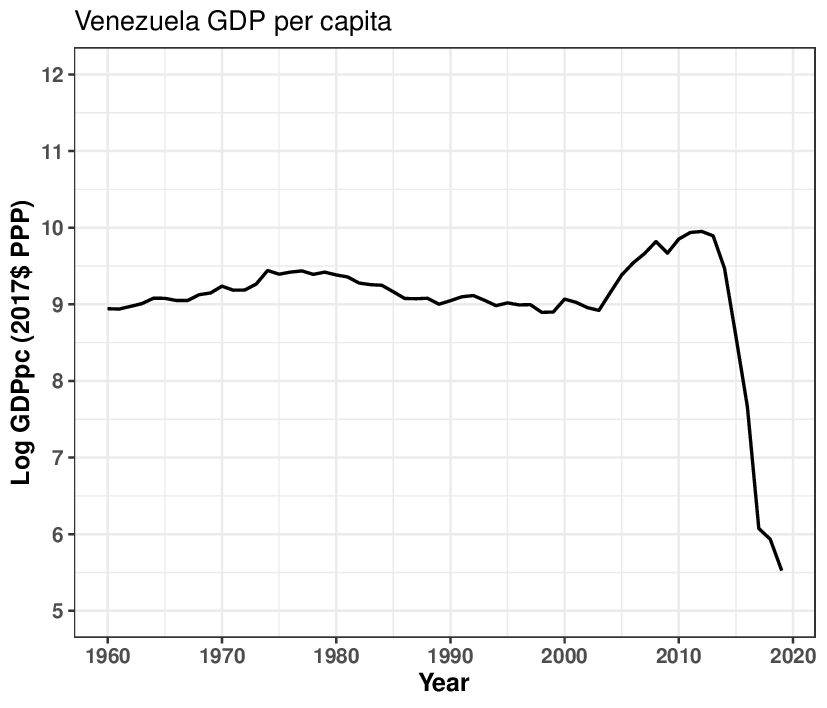}
      \caption{Log GDP per capita (2017\$ PPP) for Venezuela}
      \label{fig_ven_pc_rgdpo}
    
      \vspace{0.5ex}
      {\footnotesize
      \textbf{Notes:} The series displays a pronounced decline in Venezuela’s GDP per capita after 2010, reflecting the country’s political and economic crisis; Venezuela is treated as an outlier and excluded from the income-variance calculations.}
    \end{figure}




\end{appendices}

\bibliography{gc-bibliography}

\begin{thebibliography}{}
\renewcommand{\doi}[1]{\url{https://doi.org/#1}}
\bibcommenthead

\bibitem [\protect \citeauthoryear {%
Barro%
}{%
Barro%
}{%
{\protect \APACyear {2015}}%
}]{%
barro2015convergence}
\APACinsertmetastar {%
barro2015convergence}%
\begin{APACrefauthors}%
Barro, R.J.%
\end{APACrefauthors}%
\unskip\
\newblock
\APACrefYearMonthDay{2015}{}{}.
\newblock
{\BBOQ}\APACrefatitle {Convergence and modernisation} {Convergence and
  modernisation}.{\BBCQ}
\newblock
\APACjournalVolNumPages{The economic journal}{125}{585}{911--942,}
\newblock

\newblock

\PrintBackRefs{\CurrentBib}

\bibitem [\protect \citeauthoryear {%
Barro%
\ \BBA {} Lee%
}{%
Barro%
\ \BBA {} Lee%
}{%
{\protect \APACyear {2013}}%
}]{%
barro2013new}
\APACinsertmetastar {%
barro2013new}%
\begin{APACrefauthors}%
Barro, R.J.%
\BCBT {}\ \BBA {} Lee, J.W.%
\end{APACrefauthors}%
\unskip\
\newblock
\APACrefYearMonthDay{2013}{}{}.
\newblock
{\BBOQ}\APACrefatitle {A new data set of educational attainment in the world,
  1950--2010} {A new data set of educational attainment in the world,
  1950--2010}.{\BBCQ}
\newblock
\APACjournalVolNumPages{Journal of development economics}{104}{}{184--198,}
\newblock

\newblock

\PrintBackRefs{\CurrentBib}

\bibitem [\protect \citeauthoryear {%
Barro%
\ \BBA {} Sala-i Martin%
}{%
Barro%
\ \BBA {} Sala-i Martin%
}{%
{\protect \APACyear {1992}}%
}]{%
barro1992convergence}
\APACinsertmetastar {%
barro1992convergence}%
\begin{APACrefauthors}%
Barro, R.J.%
\BCBT {}\ \BBA {} Sala-i Martin, X.%
\end{APACrefauthors}%
\unskip\
\newblock
\APACrefYearMonthDay{1992}{}{}.
\newblock
{\BBOQ}\APACrefatitle {Convergence} {Convergence}.{\BBCQ}
\newblock
\APACjournalVolNumPages{Journal of political Economy}{100}{2}{223--251,}
\newblock

\newblock

\PrintBackRefs{\CurrentBib}

\bibitem [\protect \citeauthoryear {%
Caselli%
}{%
Caselli%
}{%
{\protect \APACyear {2005}}%
}]{%
caselli2005accounting}
\APACinsertmetastar {%
caselli2005accounting}%
\begin{APACrefauthors}%
Caselli, F.%
\end{APACrefauthors}%
\unskip\
\newblock
\APACrefYearMonthDay{2005}{}{}.
\newblock
{\BBOQ}\APACrefatitle {Accounting for cross-country income differences}
  {Accounting for cross-country income differences}.{\BBCQ}
\newblock
\APACjournalVolNumPages{Handbook of economic growth}{1}{}{679--741,}
\newblock

\newblock

\PrintBackRefs{\CurrentBib}

\bibitem [\protect \citeauthoryear {%
Cohen%
\ \BBA {} Soto%
}{%
Cohen%
\ \BBA {} Soto%
}{%
{\protect \APACyear {2007}}%
}]{%
cohen2007growth}
\APACinsertmetastar {%
cohen2007growth}%
\begin{APACrefauthors}%
Cohen, D.%
\BCBT {}\ \BBA {} Soto, M.%
\end{APACrefauthors}%
\unskip\
\newblock
\APACrefYearMonthDay{2007}{}{}.
\newblock
{\BBOQ}\APACrefatitle {Growth and human capital: good data, good results}
  {Growth and human capital: good data, good results}.{\BBCQ}
\newblock
\APACjournalVolNumPages{Journal of economic growth}{12}{}{51--76,}
\newblock

\newblock

\PrintBackRefs{\CurrentBib}

\bibitem [\protect \citeauthoryear {%
Easterly%
\ \BBA {} Levine%
}{%
Easterly%
\ \BBA {} Levine%
}{%
{\protect \APACyear {2001}}%
}]{%
easterly2001have}
\APACinsertmetastar {%
easterly2001have}%
\begin{APACrefauthors}%
Easterly, W.%
\BCBT {}\ \BBA {} Levine, R.%
\end{APACrefauthors}%
\unskip\
\newblock
\APACrefYearMonthDay{2001}{}{}.
\newblock
{\BBOQ}\APACrefatitle {What have we learned from a decade of empirical research
  on growth? It's Not Factor Accumulation: Stylized Facts and Growth Models}
  {What have we learned from a decade of empirical research on growth? it's not
  factor accumulation: Stylized facts and growth models}.{\BBCQ}
\newblock
\APACjournalVolNumPages{The world bank economic review}{15}{2}{177--219,}
\newblock

\newblock

\PrintBackRefs{\CurrentBib}

\bibitem [\protect \citeauthoryear {%
Feenstra%
, Inklaar%
\BCBL {}\ \BBA {} Timmer%
}{%
Feenstra%
\ \protect \BOthers {.}}{%
{\protect \APACyear {2015}}%
}]{%
feenstra2015next}
\APACinsertmetastar {%
feenstra2015next}%
\begin{APACrefauthors}%
Feenstra, R.C.%
, Inklaar, R.%
\BCBL {} Timmer, M.P.%
\end{APACrefauthors}%
\unskip\
\newblock
\APACrefYearMonthDay{2015}{}{}.
\newblock
{\BBOQ}\APACrefatitle {The next generation of the Penn World Table} {The next
  generation of the penn world table}.{\BBCQ}
\newblock
\APACjournalVolNumPages{American economic review}{105}{10}{3150--3182,}
\newblock

\newblock

\PrintBackRefs{\CurrentBib}

\bibitem [\protect \citeauthoryear {%
Gollin%
}{%
Gollin%
}{%
{\protect \APACyear {2002}}%
}]{%
gollin2002getting}
\APACinsertmetastar {%
gollin2002getting}%
\begin{APACrefauthors}%
Gollin, D.%
\end{APACrefauthors}%
\unskip\
\newblock
\APACrefYearMonthDay{2002}{}{}.
\newblock
{\BBOQ}\APACrefatitle {Getting income shares right} {Getting income shares
  right}.{\BBCQ}
\newblock
\APACjournalVolNumPages{Journal of political Economy}{110}{2}{458--474,}
\newblock

\newblock

\PrintBackRefs{\CurrentBib}

\bibitem [\protect \citeauthoryear {%
Hanushek%
\ \BBA {} Woessmann%
}{%
Hanushek%
\ \BBA {} Woessmann%
}{%
{\protect \APACyear {2012}}%
}]{%
hanushek2012better}
\APACinsertmetastar {%
hanushek2012better}%
\begin{APACrefauthors}%
Hanushek, E.A.%
\BCBT {}\ \BBA {} Woessmann, L.%
\end{APACrefauthors}%
\unskip\
\newblock
\APACrefYearMonthDay{2012}{}{}.
\newblock
{\BBOQ}\APACrefatitle {Do better schools lead to more growth? Cognitive skills,
  economic outcomes, and causation} {Do better schools lead to more growth?
  cognitive skills, economic outcomes, and causation}.{\BBCQ}
\newblock
\APACjournalVolNumPages{Journal of economic growth}{17}{}{267--321,}
\newblock

\newblock

\PrintBackRefs{\CurrentBib}

\bibitem [\protect \citeauthoryear {%
Johnson%
\ \BBA {} Papageorgiou%
}{%
Johnson%
\ \BBA {} Papageorgiou%
}{%
{\protect \APACyear {2020}}%
}]{%
johnson2020remains}
\APACinsertmetastar {%
johnson2020remains}%
\begin{APACrefauthors}%
Johnson, P.%
\BCBT {}\ \BBA {} Papageorgiou, C.%
\end{APACrefauthors}%
\unskip\
\newblock
\APACrefYearMonthDay{2020}{}{}.
\newblock
{\BBOQ}\APACrefatitle {What remains of cross-country convergence?} {What
  remains of cross-country convergence?}{\BBCQ}
\newblock
\APACjournalVolNumPages{Journal of Economic Literature}{58}{1}{129--175,}
\newblock

\newblock

\PrintBackRefs{\CurrentBib}

\bibitem [\protect \citeauthoryear {%
Jones%
}{%
Jones%
}{%
{\protect \APACyear {2016}}%
}]{%
jones2016facts}
\APACinsertmetastar {%
jones2016facts}%
\begin{APACrefauthors}%
Jones, C.I.%
\end{APACrefauthors}%
\unskip\
\newblock
\APACrefYearMonthDay{2016}{}{}.
\newblock
{\BBOQ}\APACrefatitle {The Facts of Economic Growth} {The facts of economic
  growth}.{\BBCQ}
\newblock
 J.B.~Taylor\ \BBA {} H.~Uhlig\ (\BEDS), \APACrefbtitle {Handbook of
  Macroeconomics} {Handbook of macroeconomics}\ (\BVOL~2, \BPGS\ 3--69).
\newblock
\APACaddressPublisher{}{Elsevier}.
\PrintBackRefs{\CurrentBib}

\bibitem [\protect \citeauthoryear {%
Karabarbounis%
\ \BBA {} Neiman%
}{%
Karabarbounis%
\ \BBA {} Neiman%
}{%
{\protect \APACyear {2014}}%
}]{%
karabarbounis2014global}
\APACinsertmetastar {%
karabarbounis2014global}%
\begin{APACrefauthors}%
Karabarbounis, L.%
\BCBT {}\ \BBA {} Neiman, B.%
\end{APACrefauthors}%
\unskip\
\newblock
\APACrefYearMonthDay{2014}{}{}.
\newblock
{\BBOQ}\APACrefatitle {The global decline of the labor share} {The global
  decline of the labor share}.{\BBCQ}
\newblock
\APACjournalVolNumPages{The Quarterly journal of economics}{129}{1}{61--103,}
\newblock

\newblock

\PrintBackRefs{\CurrentBib}

\bibitem [\protect \citeauthoryear {%
Klenow%
\ \BBA {} Rodriguez-Clare%
}{%
Klenow%
\ \BBA {} Rodriguez-Clare%
}{%
{\protect \APACyear {1997}}%
}]{%
klenow1997neoclassical}
\APACinsertmetastar {%
klenow1997neoclassical}%
\begin{APACrefauthors}%
Klenow, P.J.%
\BCBT {}\ \BBA {} Rodriguez-Clare, A.%
\end{APACrefauthors}%
\unskip\
\newblock
\APACrefYearMonthDay{1997}{}{}.
\newblock
{\BBOQ}\APACrefatitle {The neoclassical revival in growth economics: Has it
  gone too far?} {The neoclassical revival in growth economics: Has it gone too
  far?}{\BBCQ}
\newblock
\APACjournalVolNumPages{NBER macroeconomics annual}{12}{}{73--103,}
\newblock

\newblock

\PrintBackRefs{\CurrentBib}

\bibitem [\protect \citeauthoryear {%
Kremer%
, Willis%
\BCBL {}\ \BBA {} You%
}{%
Kremer%
\ \protect \BOthers {.}}{%
{\protect \APACyear {2022}}%
}]{%
kremer2022converging}
\APACinsertmetastar {%
kremer2022converging}%
\begin{APACrefauthors}%
Kremer, M.%
, Willis, J.%
\BCBL {} You, Y.%
\end{APACrefauthors}%
\unskip\
\newblock
\APACrefYearMonthDay{2022}{}{}.
\newblock
{\BBOQ}\APACrefatitle {Converging to convergence} {Converging to
  convergence}.{\BBCQ}
\newblock
\APACjournalVolNumPages{NBER macroeconomics annual}{36}{1}{337--412,}
\newblock

\newblock

\PrintBackRefs{\CurrentBib}

\bibitem [\protect \citeauthoryear {%
Lagakos%
, Moll%
, Porzio%
, Qian%
\BCBL {}\ \BBA {} Schoellman%
}{%
Lagakos%
\ \protect \BOthers {.}}{%
{\protect \APACyear {2018}}%
}]{%
lagakos2018life}
\APACinsertmetastar {%
lagakos2018life}%
\begin{APACrefauthors}%
Lagakos, D.%
, Moll, B.%
, Porzio, T.%
, Qian, N.%
\BCBL {} Schoellman, T.%
\end{APACrefauthors}%
\unskip\
\newblock
\APACrefYearMonthDay{2018}{}{}.
\newblock
{\BBOQ}\APACrefatitle {Life cycle wage growth across countries} {Life cycle
  wage growth across countries}.{\BBCQ}
\newblock
\APACjournalVolNumPages{Journal of Political Economy}{126}{2}{797--849,}
\newblock

\newblock

\PrintBackRefs{\CurrentBib}

\bibitem [\protect \citeauthoryear {%
Ndikumana%
\ \BBA {} Boyce%
}{%
Ndikumana%
\ \BBA {} Boyce%
}{%
{\protect \APACyear {2011}}%
}]{%
ndikumana2011africa}
\APACinsertmetastar {%
ndikumana2011africa}%
\begin{APACrefauthors}%
Ndikumana, L.%
\BCBT {}\ \BBA {} Boyce, J.K.%
\end{APACrefauthors}%
\unskip\
\newblock
\APACrefYear{2011}.
\newblock
\APACrefbtitle {Africa's odious debts: How foreign loans and capital flight
  bled a continent} {Africa's odious debts: How foreign loans and capital
  flight bled a continent}.
\newblock
\APACaddressPublisher{}{Bloomsbury Publishing}.
\PrintBackRefs{\CurrentBib}

\bibitem [\protect \citeauthoryear {%
Patel%
, Sandefur%
\BCBL {}\ \BBA {} Subramanian%
}{%
Patel%
\ \protect \BOthers {.}}{%
{\protect \APACyear {2021}}%
}]{%
patel2021new}
\APACinsertmetastar {%
patel2021new}%
\begin{APACrefauthors}%
Patel, D.%
, Sandefur, J.%
\BCBL {} Subramanian, A.%
\end{APACrefauthors}%
\unskip\
\newblock
\APACrefYearMonthDay{2021}{}{}.
\newblock
{\BBOQ}\APACrefatitle {The new era of unconditional convergence} {The new era
  of unconditional convergence}.{\BBCQ}
\newblock
\APACjournalVolNumPages{Journal of Development Economics}{152}{}{102687,}
\newblock

\newblock

\PrintBackRefs{\CurrentBib}

\bibitem [\protect \citeauthoryear {%
Pritchett%
}{%
Pritchett%
}{%
{\protect \APACyear {1997}}%
}]{%
pritchett1997divergence}
\APACinsertmetastar {%
pritchett1997divergence}%
\begin{APACrefauthors}%
Pritchett, L.%
\end{APACrefauthors}%
\unskip\
\newblock
\APACrefYearMonthDay{1997}{}{}.
\newblock
{\BBOQ}\APACrefatitle {Divergence, big time} {Divergence, big time}.{\BBCQ}
\newblock
\APACjournalVolNumPages{Journal of Economic perspectives}{11}{3}{3--17,}
\newblock

\newblock

\PrintBackRefs{\CurrentBib}

\bibitem [\protect \citeauthoryear {%
Psacharopoulos%
}{%
Psacharopoulos%
}{%
{\protect \APACyear {1994}}%
}]{%
psacharopoulos1994returns}
\APACinsertmetastar {%
psacharopoulos1994returns}%
\begin{APACrefauthors}%
Psacharopoulos, G.%
\end{APACrefauthors}%
\unskip\
\newblock
\APACrefYearMonthDay{1994}{}{}.
\newblock
{\BBOQ}\APACrefatitle {Returns to investment in education: A global update}
  {Returns to investment in education: A global update}.{\BBCQ}
\newblock
\APACjournalVolNumPages{World development}{22}{9}{1325--1343,}
\newblock

\newblock

\PrintBackRefs{\CurrentBib}

\bibitem [\protect \citeauthoryear {%
Schoellman%
}{%
Schoellman%
}{%
{\protect \APACyear {2012}}%
}]{%
schoellman2012education}
\APACinsertmetastar {%
schoellman2012education}%
\begin{APACrefauthors}%
Schoellman, T.%
\end{APACrefauthors}%
\unskip\
\newblock
\APACrefYearMonthDay{2012}{}{}.
\newblock
{\BBOQ}\APACrefatitle {Education quality and development accounting} {Education
  quality and development accounting}.{\BBCQ}
\newblock
\APACjournalVolNumPages{The Review of Economic Studies}{79}{1}{388--417,}
\newblock

\newblock

\PrintBackRefs{\CurrentBib}

\bibitem [\protect \citeauthoryear {%
A.~Young%
}{%
A.~Young%
}{%
{\protect \APACyear {2012}}%
}]{%
young2012african}
\APACinsertmetastar {%
young2012african}%
\begin{APACrefauthors}%
Young, A.%
\end{APACrefauthors}%
\unskip\
\newblock
\APACrefYearMonthDay{2012}{}{}.
\newblock
{\BBOQ}\APACrefatitle {The African growth miracle} {The african growth
  miracle}.{\BBCQ}
\newblock
\APACjournalVolNumPages{Journal of Political Economy}{120}{4}{696--739,}
\newblock

\newblock

\PrintBackRefs{\CurrentBib}

\bibitem [\protect \citeauthoryear {%
A.T.~Young%
, Higgins%
\BCBL {}\ \BBA {} Levy%
}{%
A.T.~Young%
\ \protect \BOthers {.}}{%
{\protect \APACyear {2008}}%
}]{%
young2008sigma}
\APACinsertmetastar {%
young2008sigma}%
\begin{APACrefauthors}%
Young, A.T.%
, Higgins, M.J.%
\BCBL {} Levy, D.%
\end{APACrefauthors}%
\unskip\
\newblock
\APACrefYearMonthDay{2008}{}{}.
\newblock
{\BBOQ}\APACrefatitle {Sigma convergence versus beta convergence: Evidence from
  US county-level data} {Sigma convergence versus beta convergence: Evidence
  from us county-level data}.{\BBCQ}
\newblock
\APACjournalVolNumPages{Journal of Money, Credit and
  Banking}{40}{5}{1083--1093,}
\newblock

\newblock

\PrintBackRefs{\CurrentBib}

\end{thebibliography}

\end{document}